\documentclass[prb,twocolumn,amsmath,amssymb,superscriptaddress,floatfix]{revtex4}
\usepackage[english]{babel}
\usepackage{amsfonts}
\usepackage{graphicx}
\usepackage{times}
\usepackage{braket}
\usepackage{soul}
\usepackage{subcaption}

\usepackage{color}
\definecolor{nred} {RGB}{224,0,0}
\definecolor{nblue} {RGB}{28,130,185}
\definecolor{dgreen} {RGB}{78,138,21}

\begin{document}

\title{Spectral function of a bipolaron coupled to dispersive optical phonons}

\author{K. \surname{Kova\v c} }
\affiliation{J. Stefan Institute, 1000 Ljubljana, Slovenia}
\affiliation{Faculty of Mathematics and Physics, University of Ljubljana, 1000
	Ljubljana, Slovenia}

\author{J. \surname{Bon\v ca}}
\affiliation{J. Stefan Institute, 1000 Ljubljana, Slovenia}
\affiliation{Faculty of Mathematics and Physics, University of Ljubljana, 1000
	Ljubljana, Slovenia}

\date{\today}
\begin{abstract}
Using an efficient variational exact diagonalization method, we computed the electron removal spectral function within the framework of the Holstein-Hubbard model containing   two electrons with opposite spins  coupled to  dispersive quantum optical phonons. Our primary focus was examining the interplay between phonon dispersion and Coulomb repulsion and their effects on the single--electron removal spectral function, relevant for the analysis of angle--resolved photoemission spectroscopy (ARPES). Tuning the strengths of the electron-phonon coupling and the Hubbard interaction allows us to examine the evolution of the spectral properties of the system as it crosses over from a bound bipolaron to separate polarons. With increasing Hubbard repulsion, the decrease of the bipolaron binding energy results in the gradual downward shift of the polaron band - a low-frequency feature in the spectral function. Simultaneously, the intensity of the polaron band away from the center of the Brillouin zone diminishes until it remains non-zero only in its center as the bipolaron unbinds into two separate polarons. The spectral function is significantly influenced by phonon dispersion, particularly in systems with strong electron-phonon coupling. The sign of the curvature of the phonon band plays a crucial role in the distribution of spectral weight. 
\end{abstract}  

\maketitle

\setcounter{figure}{0}


\section{Introduction}

The interaction between charge carriers and quantum lattice vibrations, known as electron-phonon (EP) coupling, has a profound effect on material properties \cite{Lanzara1998, Miyata_2017, Huang_2019, Cinquanta2019, Ghosh2020, Karmakar2022}. In recent years, EP coupling has garnered considerable attention in the study of organic semiconductors \cite{Mladenović2014, Fratini2020, Chang2022}, where it strongly influences electronic structure and charge transport properties. The Holstein model (HM), which represents a molecular crystal where electrons interact locally with intramolecular vibrations, is a fundamental model for studying these interactions \cite{Holstein1959}. Extensive research has investigated the HM using both analytical and numerical methods across various regimes \cite{Ranninger1992, Marsiglio1993, Alexandrov1994, Wellein1997, Fehske_1997, Capone1997, Wellein1998, Bonca1999, Fehske2000, Ku2002, Barisic2002, Hohenadler2003, Berciu2006, Goodvin2006, Fehske2007, Barisic2012, Adolphs2014, Adolphs2014_2, Chandler2016, Yam2020, marsiglio2022, Mitric2022, Mitric2023, Nocera2023, Zhao2023}. \\
The HM simplifies EP interactions by focusing on short-range interactions between charge carriers and lattice distortions, assuming effective screening of long-range interactions. The strength of this coupling depends on relative displacements between the charge carrier atom and its neighbors. While most studies have focused on short-range coupling to optical phonons, there have been fewer investigations into short-range coupling with acoustic phonons due to their negligible relative displacements \cite{Li2011, Hahn2021, Li_2013}.
Optical phonons, involving antiphase atomic motion with substantial relative displacements, lead to stronger electron-phonon coupling and are commonly included in the HM as dispersionless Einstein phonons. This choice is motivated firstly by its simplicity in analytical and numerical treatments and secondly by being a reasonable approximation in the case when the phonon bandwidth is small compared to the average phonon frequency. The existing body of research studying the HM with dispersive optical phonons is relatively limited \cite{Marchand2013, Bonca2021, Bonca2022, Jansen2022, Kovac2024}. These studies have demonstrated that phonon dispersion exerts a profound influence on polaron properties, impacting quantities such as effective mass \cite{Marchand2013}, optical conductivity \cite{Jansen2022}, and spectral function \cite{Bonca2021, Bonca2022}. However, it's important to note that these investigations primarily focus on how dispersion affects polaron properties and do not extensively explore the interactions between polarons. In our previous study \cite{Kovac2024}, we investigated the effect of phonon dispersion in a system with two electrons coupled to optical phonons within the Holstein-Hubbard model (HHM), which incorporates Coulomb interaction between charge carriers. We found that phonon dispersion significantly influences the stability of bound states, particularly in systems where the phonon cloud spreads over multiple sites due to sufficient Coulomb interaction. The HHM can yield in the low-doping limit various types of bipolarons based on its parameters, as evidenced in prior studies \cite{Bonca2000, Bonca2001}. In regimes dominated by EP coupling, both polarons tend to localize on the same lattice site, forming an S0 bipolaron. However, with increasing Coulomb repulsion strength, double occupancy becomes less favorable.
Consequently, the bipolaron spreads across multiple lattice sites, and the probability distribution of electron occupation determines its specific type. For instance, in an S1 bipolaron, electrons exhibit a maximal probability of occupying neighboring lattice sites. Similarly, for an S2 bipolaron, maximal probability extends to next-nearest neighboring sites, and so forth. 
\\
Recent advancements in angle-resolved photoemission spectroscopy (ARPES) have made it a powerful tool for studying electronic states in interacting systems~\cite{Damascelli2004, Comin2014, Kordyuk2015, Boschini2024}. Experimental efforts using ARPES to explore EP-coupled systems highlight the need for theoretical insights to interpret experimental data~\cite{Shen2007, Cancellieri2016, Strocov2018, Kang2018, Krsnik2020, Sajedi2022}. In this study, we aim to investigate how phonon dispersion affects spectral functions, which are directly related to ARPES spectra. We aim to identify and analyze transitions between different bipolaronic regimes by observing spectral functions. It's worth noting that our computational approach is constrained by the exponential growth of the Hilbert space, limiting our investigation to systems with a low carrier doping level ($n_{dop}\rightarrow 0$).  Recently ARPES data has been published for various systems exhibiting low carrier doping and showcasing notable polaronic effects \cite{Verdi2017, Caruso2018, Chen2018}. In recent work~\cite{Kovac2024a}, authors present a rather unexpected similarity between the spectral functions of HHM at low doping levels and those obtained based on an approximation that the low-doping system is composed of bipolarons that behave as hard-core bosons. This finding underscores the importance of investigating low-doped systems, highlighting their relevance and providing a solid basis for our study.  \\
The structure of this paper is organized as follows. In the next section, we introduce the Hamiltonian and provide a detailed description of its formulation. We also outline the numerical methods used for constructing the basis and calculating the spectral functions. Section III presents the core results of our work, starting with the spectra of one- and two-electron systems, which are crucial for understanding the spectral functions. The primary focus is then placed on the spectral functions themselves and their detailed analysis. Finally, we conclude the paper with a summary of our findings and key insights.


\section{Model and method}

We consider a system of two electrons coupled to dispersive optical phonons described by the Hamiltonian

\begin{eqnarray} \label{hol}
	\vspace*{-0.0cm}
	\mathcal{H}_0 &=& -t_\mathrm{el}\sum_{j,s}(c^\dagger_{j,s} c_{j+1,s} +\mathrm{H.c.}) + {g} \sum_{j} \hat n_{j}(b_{j}^\dagger + b_{j})+ \nonumber \\
	&+& \omega_0\sum_{j} b_{j}^\dagger  b_{j} + t_\mathrm{ph}\sum_{j}(b^\dagger_{j} b_{j+1} +\mathrm{H.c.}) + \nonumber \\ &+& U\sum_{j}\hat n_{j,\uparrow}\hat n_{j,\downarrow}, 
	\label{ham}
\end{eqnarray}

In this Hamiltonian, $c^\dagger_{j,s}$ and $b^\dagger_{j}$ are electron and phonon creation operators at site $j$ and spin $s$, respectively. The operator $\hat n_{j} = \sum_s c^\dagger_{j,s}c_{j,s}$ represents the electron density at site $j$, and spin $s$.  From here on we set $t_\mathrm{el}$, the nearest-neighbor electron hopping amplitude, to $1$ . \\
The optical phonon band is represented as $\omega_\mathrm{ph}(q) = \omega_0 + 2t_\mathrm{ph}\cos(q)$, characterized by two parameters: $\omega_0$ (determining the central position of the band) and $t_\mathrm{ph}$ (governing its bandwidth). Notably, when $t_\mathrm{ph}<0$, the signs of the phonon and electron dispersion curvatures overlap. From here on we set $\omega_0$ to 1. \\
The second term in  Eq.~(\ref{ham}) describes the interaction between electrons and phonons, and the last term is the on-site Coulomb repulsion. In the study of HM, it is customary to introduce a dimensionless parameter that characterizes the system, namely the effective EP coupling strength $\lambda$ defined as $\lambda = g^2/2t_\mathrm{el}\sqrt{\omega_0^2 - 4t_\mathrm{ph}^2}$. In this study, we focus on how phonon dispersion influences system behavior, expressing our results in terms of the EP coupling $g$ rather than $\lambda$.  \\
We employed a numerical method described in detail in Refs. \cite{Bonca1999,Ku2002,Bonca2021}. A variational subspace is constructed iteratively beginning with a particular state where both electrons are on the same site with no phonons. The subspace is initially defined on an infinite one-dimensional lattice.   The variational Hilbert space is then generated by applying a sum of two off--diagonal operators:$\sum_{j,s}(c^\dagger_{j,s} c_{j+1,s} +\mathrm{H.c.}) + \sum_{j} \hat n_{j} (b_{j}^\dagger + b_{j})$, $N_h$ times taking into account the full translational symmetry. The obtained subspace is restricted in the sense that it allows only a finite maximal distance of a phonon quanta from the doubly occupied site, $l_\mathrm{max_1}=(N_\mathrm{h}-1)/2$, a maximal distance between two electrons $l_\mathrm{max_2}=N_\mathrm{h}$, and a maximal amount of phonon quanta at the doubly occupied site $N_\mathrm{phmax}(0)=N_\mathrm{h}$, while on the site, which is $l$ sites away from the doubly occupied one, it is reduced to $N_\mathrm{phmax}(l)=N_\mathrm{h} - 2l-1$. 
The procedure generates variational Hilbert space  with   $L=2N_h + 1$ sites, and  periodic boundary conditions. 
We have used a standard Lanczos procedure \cite{Lanczos1950} to obtain static properties of the model. If not otherwise specified $N_h$ is set to $16$.  \\
The main objective of this work is to analyze the spectral function of the two-electron system corresponding to the removal of an electron with spin $s$ at $T = 0$. The spectral function $A(\omega,k)$ is defined as 

\begin{multline} \label{spektralka}
    A(\omega,k) = \sum_{j = 0}^{M}\Big{|}\bra{\psi_{-k}^{(j,1)}}c_{ks}\ket{\psi_{0}^{(0,2)}}\Big{|}^2\\
    \times\delta\Big{(}\omega-E_{-k}^{(j,1)}+ E_0^{(0,2)}\Big{)},
\end{multline}
where $\ket{\psi_{k}^{(n,N)}}$ represents the n-th translationally invariant state with $N$ electrons and wave vector $k$. We computed $M = 2000$ one-electron states $\ket{\psi_{k}^{(n,1)}}$ for each $k$, ensuring orthogonality via the Gram-Schmidt reorthogonalization procedure. We used a Lorentzian form of $\delta$-functions with a half-width at half-maximum of $\eta = 0.05$ for graphic representations of $A(\omega,k)$.\\
To facilitate the presentation of density plots later in this paper, we extended our calculations beyond conventional periodic boundary conditions to incorporate twisted boundary conditions, as discussed in works such as \cite{Shastry1990, Bonca2003}. Twisted boundary conditions represent a magnetic flux penetrating the ring structure, enabling a continuous connection between discrete $k$ points. Under these conditions, the kinetic energy term in Eq.~\eqref{hol} is adjusted as follows

\begin{equation}
    \mathcal{H}_{kin} = -t_\mathrm{el}\sum_{j,s}(c^\dagger_{j,s} c_{j+1,s}e^{i\theta} +\mathrm{H.c.}).
\end{equation}

For a system of free electrons, this transformation leads to the dispersion relation $\varepsilon(k,\theta) = -2t_{el} \cos(k + \theta)$, where $\theta$ is the Peierls phase, defined as $\theta = \frac{e_0 \phi_m}{\hbar L}$. This phase arises due to the magnetic flux $\phi_m$ permeating the ring. 
Introducing a magnetic flux enables us to compute the physical properties of the system, such as the spectral function, on a more dense set in the k-space. 
Without magnetic flux $k$-vectors  are restricted to $k_n = 2\pi n/L$ within the interval  $[-\pi, \pi]$. By employing twisted boundary conditions with an appropriate choice of a set of $\theta$, additional $\mathcal{S} - 1$ points can be introduced between two neighboring $k_n$, effectively expanding the $k$-point sampling to $k_n = \frac{2\pi n}{\mathcal{S}L}$. This results in an increase in the total number of $k$-vectors to $L\mathcal{S}$, where $\mathcal{S}$ denotes the scaling factor. The spectral function we present is effectively a sum of spectral functions obtained for different $\theta$-values; therefore, it should be rescaled accordingly. As a note of caution, we stress that while 
the proposed approach using twisted boundary conditions is an exact
in the case of free electrons, in the case of interacting many
particle systems it represent an approximation. The advantage
of this approach and its impact on the spectral function are
presented in Appendix B.\\
To validate our spectral function computation, we assessed its consistency with theoretical expectations using a sum rule analysis. We evaluate the integral of $A(\omega, k)$ over the entire frequency range

\begin{equation}    \label{SumRule}
    \int_{-\infty}^{\infty} A(\omega, k) \, d\omega = \bra{\psi_{0}^{(0,2)}} c_{ks}^\dagger c_{ks} \ket{\psi_{0}^{(0,2)}} = n_{ks}
\end{equation}

Here, $n_{ks}$ represents the electron occupation number for the wavevector $k$ and spin $s$, assuming a complete basis of one-electron states is used. 
The sum rule - summation  of $n_{ks}$ across the entire Brillouin zone in the case of a system with two electrons with opposite spins  - gives 


\begin{equation}\label{SumRule1}
    \sum_{k}n_{ks} = 1
\end{equation}

Due to the expanded $k$-point sampling, the summation of the spectral function over the entire frequency range yields $\mathcal{S}n_{ks}$. To satisfy the sum rule (Eq.~\eqref{SumRule1}), we rescale the spectral function as $A(\omega, k)/\mathcal{S}$. From this point onward, we set $\mathcal{S} = 3$, as this choice ensures a visibly smooth spectral function, as demonstrated in Appendix B.\\
It is important to note that while the Hilbert space of the Hamiltonian is infinite, the numerical basis we use is finite and constructed from states that are most relevant for capturing the physics of the model. Because we do not use the complete basis, the sum rule in Eq.~\eqref{SumRule1} should be modified as 
\begin{equation}
   I= \sum_{k}\int_{-\infty}^{\infty}A(\omega,k)\mathrm{d}\omega \leq \sum_{k}\bra{\psi_{0}^{(0,2)}}n_{ks}\ket{\psi_{0}^{(0,2)}} = 1,
\end{equation}
and serves as a test of our numerical approach. By choosing sufficiently large $M$, we obtain   $I\gtrsim  0.98$.


\section{Results}

\begin{figure}[!tbh]
	\begin{subfigure}[b]{1\columnwidth}
		\centering
		\includegraphics[width=1\linewidth]{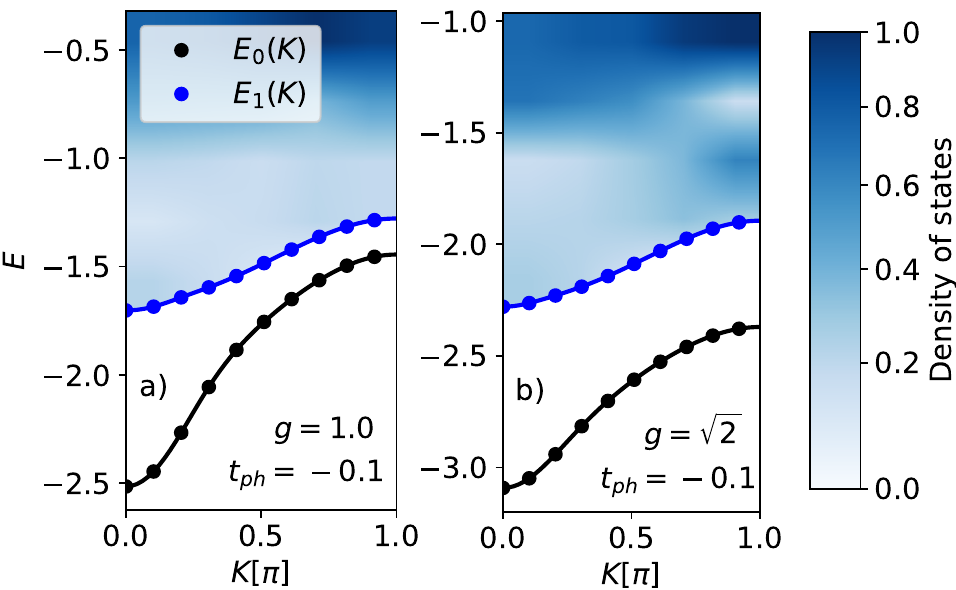} 
		\phantomsubcaption{}
	\end{subfigure}

    \vspace{-0.2cm}
    
    \begin{subfigure}[b]{1\columnwidth}
		\centering
		\includegraphics[width=1\linewidth]{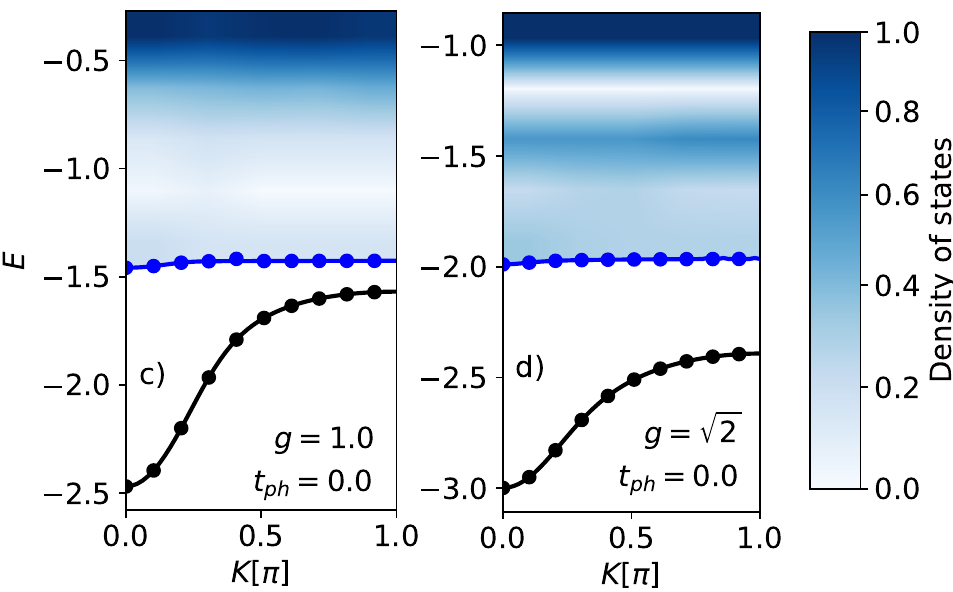} 
		\phantomsubcaption{}
	\end{subfigure}

    \vspace{-0.2cm}
      
    \begin{subfigure}[b]{1\columnwidth}
		\centering
		\includegraphics[width=1\linewidth]{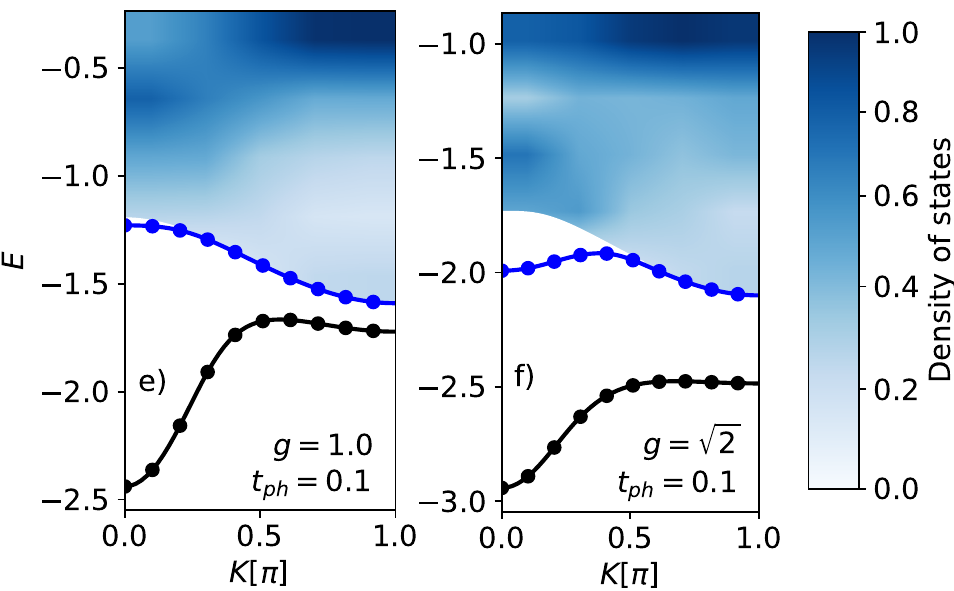} 
		\phantomsubcaption{}
		\label{fig1e}
	\end{subfigure}

    \vspace{-0.5cm}

	\captionsetup{justification=raggedright,singlelinecheck=false}
	\caption{Energy spectrum of the one-electron system is illustrated. The left column corresponds to the intermediate coupling regime, while the right column represents the strong coupling regime. Our observations reveal the spectrum's sensitivity to phonon dispersion: the middle row features figures generated for the system with Einstein phonons, the upper row depicts upward dispersion, and the bottom row represents downward dispersion. The black line corresponds to the ground state, the blue line represents the first excited state, and a continuum of states above the first excited state is depicted by a blue region with varying intensity, indicating the density of states within those regions. To enhance visualization, only every 8th data point is shown on the plot.} 
\label{fig1}
\end{figure}

In the following, we investigate two EP coupling regimes: an intermediate regime where $g$ is set to 1, and a strong coupling regime with $g = \sqrt{2}$. To better grasp the spectral function, we begin by studying the energy spectra. We focus just on the essential aspects for analyzing spectral functions, given that the spectrum has been extensively explored across numerous papers, including those incorporating phonon dispersion \cite{Bonca2019}.  At  $T = 0$,  the ground state of the two-electron system has a total $K=0$.  For the analysis of spectral functions, it is advantageous to present the one electron spectra presented in Fig.~\ref{fig1} for two distinct EP coupling constants, $g = 1$ and $\sqrt{2}$, along with three variations in phonon dispersions, $t_{ph} = -0.1$, $0$, and $0.1$. The black line in the spectrum corresponds to the ground state, while the blue line represents the first excited state. Additionally, a continuum of states above the first excited state is depicted by a blue region with varying intensity, indicating the density of states within the continuum. Throughout the entire Brillouin zone, there exists a separation between the ground state and the continuum. Notably, as observed in Figs.~\ref{fig1}e and \ref{fig1}f for $t_{\mathrm{ph}} = 0.1$, there is also a gap between the first excited state and the continuum, which diminishes as one moves away from the center of the Brillouin zone. A comparison of the two figures reveals that this separation is more pronounced in the system with a larger value of $g$. This state corresponds to the so-called bound polaron state, characteristic of the strong-coupling regime, comprising an excited polaron: a polaron with an integrated additional phonon excitation~\cite{Bonca1999, Barisic2004, Bonca2019}. Additionally, we observe that with stronger coupling, the bandwidth of the ground state narrows. 
Shifting our attention to the continuum of states resulting from various momentum distributions between electron and phonons, we note a region of vanishing density of states within the continuum in the strong coupling regime. In the case of dispersionless phonons (depicted in Fig.~\ref{fig1}d), this region persists throughout the entire Brillouin zone. However, when considering phonon dispersion, it remains detectable only in certain parts of the Brillouin zone.



\begin{figure}[!tbh]
	\begin{subfigure}[b]{0.9\columnwidth}
		\centering
		\includegraphics[width=1\linewidth]{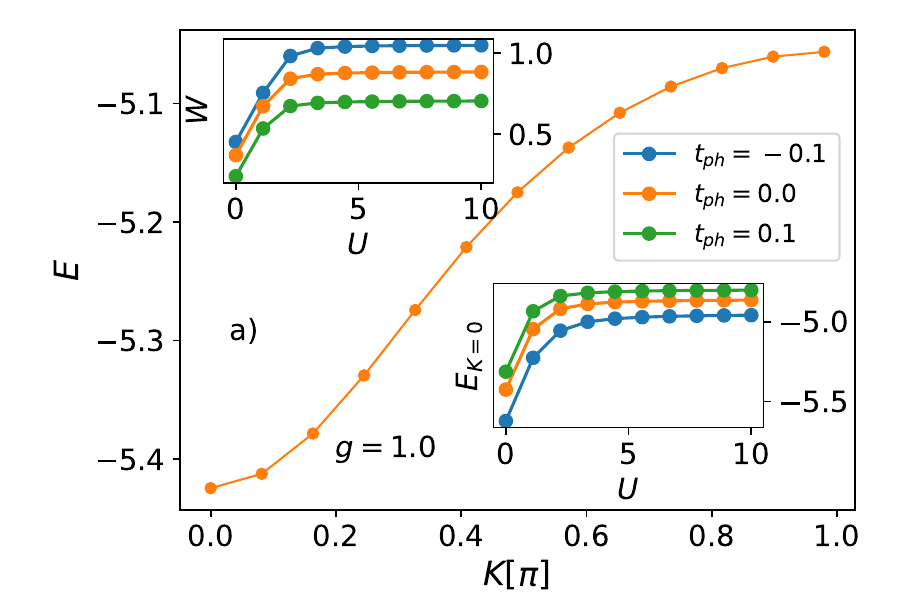} 
		\phantomsubcaption{}
		\label{fig2a}
	\end{subfigure}
 
	\begin{subfigure}[b]{0.9\columnwidth}
		\centering
		\includegraphics[width=1\linewidth]{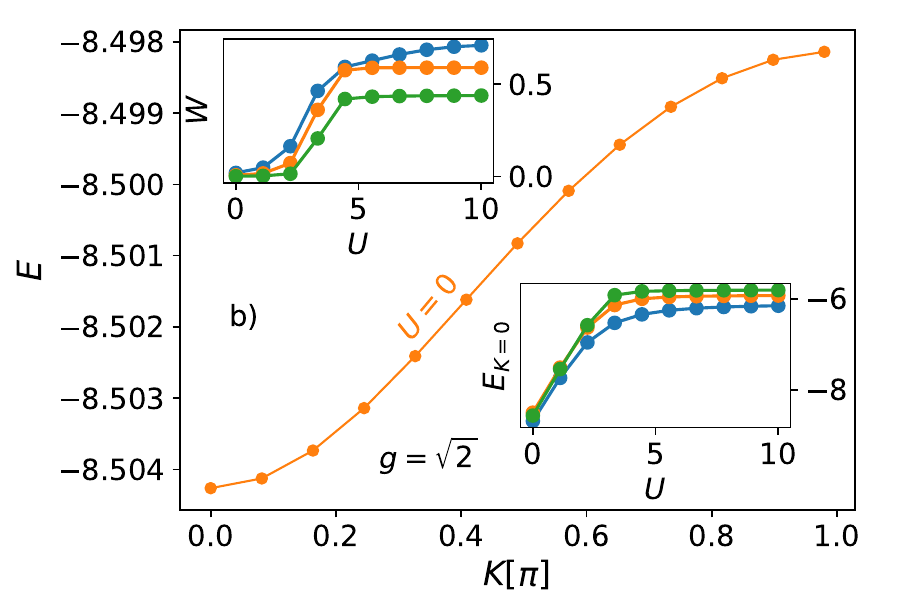} 
		\phantomsubcaption{}
		\label{fig2b}
	\end{subfigure}

	\captionsetup{justification=raggedright,singlelinecheck=false}
	\caption{The central sections of the figures illustrate the energy dispersion of the ground state for a two-electron system with $g = 1$ (Fig.~\ref{fig2a}) and $g = \sqrt{2}$ (Fig.~\ref{fig2b}), both at $U = 0$ and $t_\mathrm{ph}=0$. In the upper-left insets, we depict the variation of bandwidth ($W$) with increasing electron-electron interaction ($U$). Conversely, the lower-right insets illustrate the shift of the band with $U$. The influence of $t_\mathrm{ph}$ is showcased in these insets. To enhance visualization, only every 8th data point is shown on the plot.}
\label{fig2}
\end{figure}

 Fig.~\ref{fig2} presents the energy dispersion $E(K)$ of the ground state for a two-electron system. The main portion of the figure displays the energy dispersion for a system where electron-electron interaction and phonon dispersion are disregarded. Insets provide insights into how the bandwidth and the position of the band vary with $U$ and their reliance on phonon dispersion. Comparing with Fig.~\ref{fig1} we observe that bands are much flatter and their energy much lower in a two-electron system if Coulomb interaction is weak. This is consistent with the strong coupling limit where the S0 bipolaron effective mass is approximately  $\exp[3({g/\omega_0})^3]$-times larger than the polaron one\cite{Bonca2000}. As $U$ increases, the bandwidth $W$ expands, and the center of the band shifts towards twice the value of the ground state energy of the one-electron system. 

\begin{figure}[!tbh]
	\includegraphics[width=0.9\columnwidth]{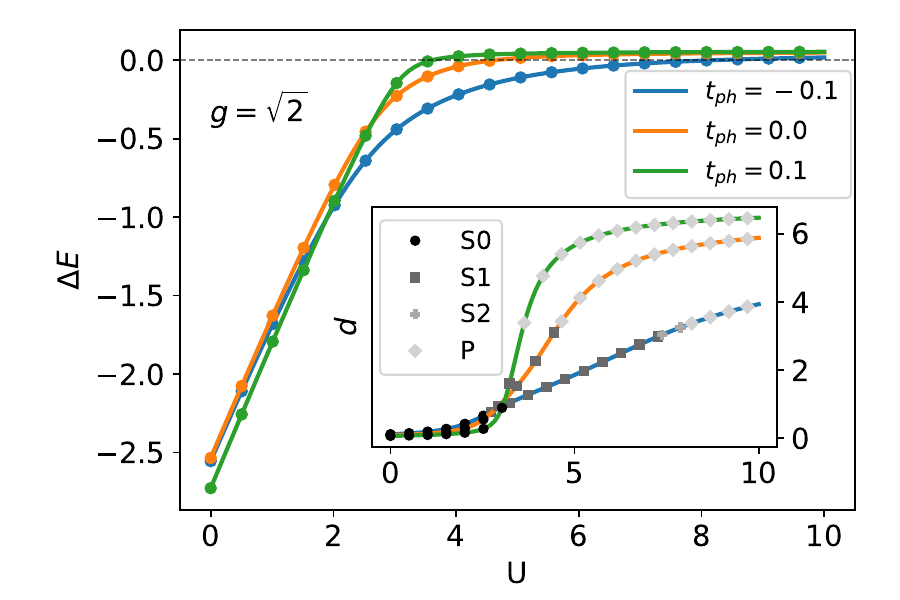}
	\captionsetup{justification=raggedright,singlelinecheck=false}
	\caption{The binding energy $\Delta E$ is plotted as a function of $U$ for three different values of $t_{\mathrm{ph}}$ in a system with $g=\sqrt{2}$. The inset displays the corresponding distance between polarons, with additional markers illustrating different regimes (bipolaronic - S0, S1, S2, polaronic - P). To enhance visualization, only every 8th data point is shown on the plot.}
	\label{fig3}
\end{figure}

We define the binding energy as $\Delta E = E_2 - 2E_1$, where $E_1$ and $E_2$ are the ground state energies of the one-electron and two-electron systems, respectively. Fig.~\ref{fig3} illustrates the dependence of $\Delta E$ on $U$ for systems with different phonon dispersions at $g=\sqrt{2}$. It can be observed that as $U$ increases, the binding energy gradually decreases towards zero, but due to the finite size of our system, it becomes slightly positive. This phenomenon arises because at $U\rightarrow\infty$, the polarons would ideally move infinitely apart from each other ($d\rightarrow\infty$) if the system were infinite. However, in our finite system, as depicted in the inset of Fig.~\ref{fig3}, $d$ converges to a finite value that is much smaller than the size of our system, its upper limit given by  $L_{max2}$. We calculate the distance as $d = \sum_j |j| p(j)$, where $j$ represents the distance between electrons with opposite spins, and $p(j)$ denotes the electron density-density correlation function defined as $p(j) = \langle \sum_i n_{i,\uparrow} n_{i+j,\downarrow} \rangle$, where $\langle \rangle$ represents the expected value in the ground state of the two-electron system. In the inset of Fig.~\ref{fig3}, additional markers are plotted along lines representing the dependence of $d$ on $U$. These markers indicate the regime in which our system resides, distinguishing between different bipolaronic types (S0, S1, S2) or unbound (P-polaronic) states at specific values of $U$. As demonstrated in our previous paper \cite{Kovac2024}, the value of $t_\mathrm{ph}$ plays a crucial role in determining the stability of the bipolaronic state. In particular, we have shown that when the sign of the phonon dispersion curvature matches that of the electron dispersion curvature, the bipolaron remains bound in the strong coupling limit, even as $U\rightarrow\infty$. In contrast, at moderate electron-phonon coupling, a light bipolaron persists up to large values of $U$. This binding arises from the exchange of phonons between two electrons residing on nearby sites, a process that is not possible without phonon dispersion. The sign of the dispersion determines whether such processes will raise or lower the energy of the polaronic pair.\cite{Kovac2024}

\begin{figure}[!tbh]
    \hspace*{-0.1\linewidth}
	\begin{subfigure}[b]{1\columnwidth}
		\centering
		\includegraphics[width=1.05\linewidth]{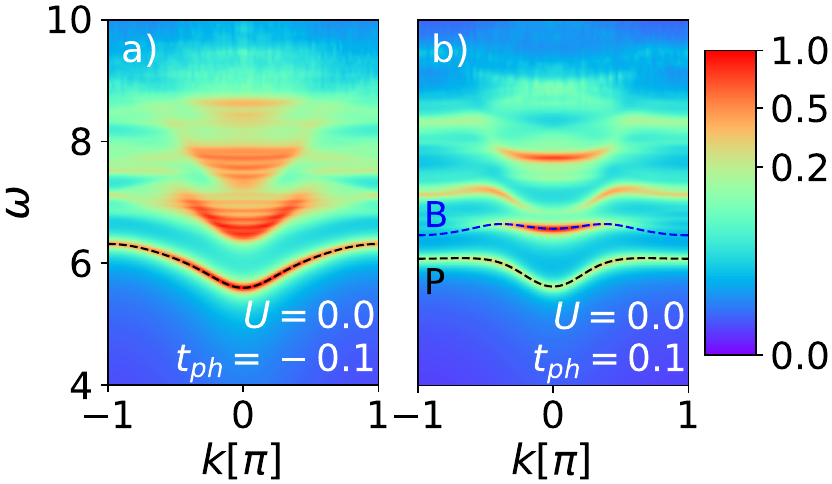} 
		\phantomsubcaption{}
		\label{fig4a}
	\end{subfigure}

    \vspace{-0.3cm}

    \begin{subfigure}[b]{1\columnwidth}
		\centering
		\includegraphics[width=1\linewidth]{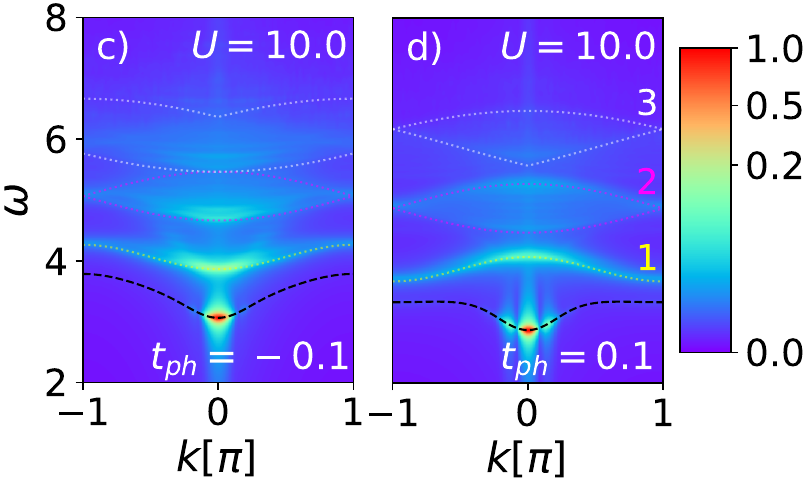} 
		\phantomsubcaption{}
		\label{fig4c}
	\end{subfigure}

    \vspace{-0.3cm}
 
	\captionsetup{justification=raggedright,singlelinecheck=false}
	\caption{The spectral functions $A(\omega,k)$, defined by Eq.~\eqref{spektralka}, are depicted. The calculations are performed in the near-strong coupling regime ($g = \sqrt{2}$). The top row corresponds to $U = 0$, while the bottom row represents the regime where the electron-electron interaction is sufficient for the bipolaron to unbind ($U = 10$). These figures illustrate the sensitivity of the spectral function to phonon dispersion: the left column depicts upward dispersion, while the right column represents downward dispersion. In these plots, the black dots correspond to the polaronic band (P), while the blue line represents the so-called bound polaron state (B). Dashed lines of different colors indicate single- (1), two- (2), and three-phonon (3) excitations, determined by Eq. (in the article). We have employed an artificial broadening parameter $\eta = 0.05$ for the illustration of delta functions. The normalization procedure is explained in the text.}
\label{fig4}
\end{figure} 

\begin{figure}[!tbh]
	\includegraphics[width=0.9\columnwidth]{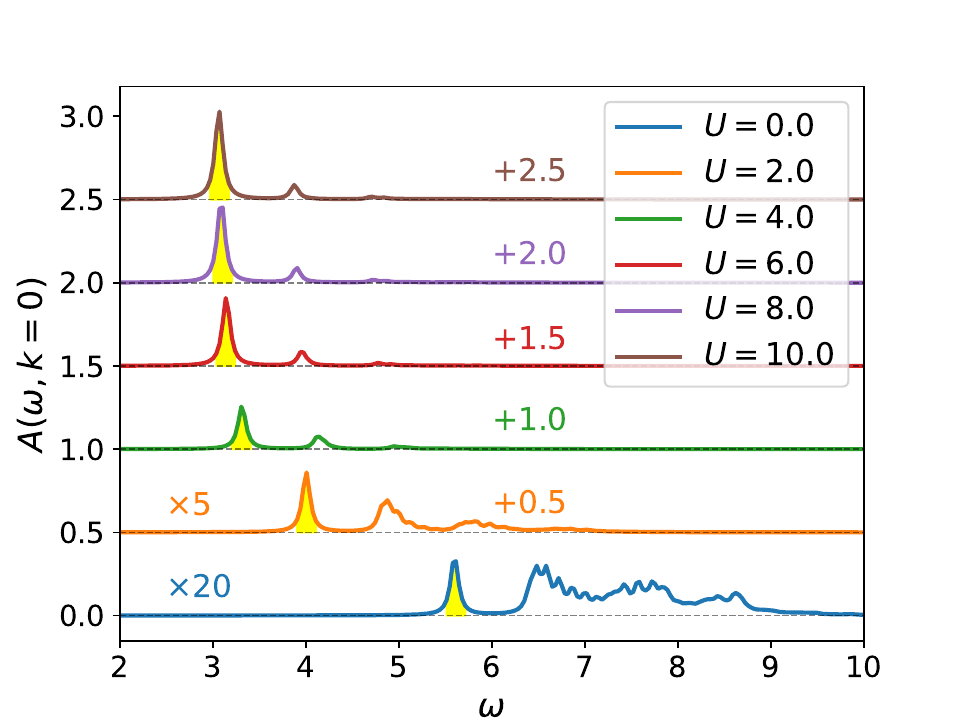}
	\captionsetup{justification=raggedright,singlelinecheck=false}
	\caption{The spectral functions at $k = 0$ in a near strong coupling regime ($g = \sqrt{2}$) with upward phonon dispersion ($t_{\mathrm{ph}} = -0.1$) are displayed at different values of $U$. To visualize them on the same figure, we applied a displacement in multiples of one-half, indicated above each line with a number of the same color. In a system where electron-electron interaction is weak, the spectral function is spread over a larger portion of the energy interval, resulting in lower spectral weight for its peaks. Therefore, for clarity, we appropriately multiplied $A(\omega,k=0, U = 0.0)$ and $A(\omega,k=0, U = 2.0)$ by a multiplier displayed above the corresponding line in the left part of the figure. The yellow region around the maximum of each spectral function illustrates the integration region used to calculate $\mathcal{W}$ for normalizing the spectral functions.}
	\label{fig5}
\end{figure}

After examining the spectra of our system of interest, we proceed with the analysis of the spectral functions. We focus on a system in a near-strong coupling regime, $g = \sqrt{2}$, as it displays much more diverse behavior - different types of bipolarons - compared with the intermediate regime. Fig.~\ref{fig4} illustrates the spectral functions for different phonon dispersions with $U = 0$ and $U = 10$. The values of Hubbard's parameter are chosen such that in the first case, there is no Coulomb interaction between the electrons, and in the second case, the interaction is sufficient to prevent the binding of polarons into a bipolaron. Spectral functions are normalized by introducing the following weight:

\begin{equation}
    \mathcal{W} = \int_{\omega_{max} - \epsilon}^{\omega_{max} + \epsilon}A(\omega,k=0)\mathrm{d}\omega,
\end{equation}
where $\omega_{max}$ denotes the position of the spectral function maximum. We calculate the one-dimensional integral by taking a slice of the spectral function at $k = 0$, as displayed in Fig.~\ref{fig5} where integration regions determined by $\epsilon$ are illustrated with a yellow area under the highest peak of $\smash{A(\omega,k=0)}$. Throughout the paper, normalized spectral functions, $A_{\text{norm}}(\omega,k) = A(\omega,k) / \mathcal{W}$, are displayed, but we omit the subscript \textit{norm}. 

\begin{figure}[!tbh]
	\begin{subfigure}[b]{0.5\columnwidth}
		\centering
		\includegraphics[width=1\linewidth]{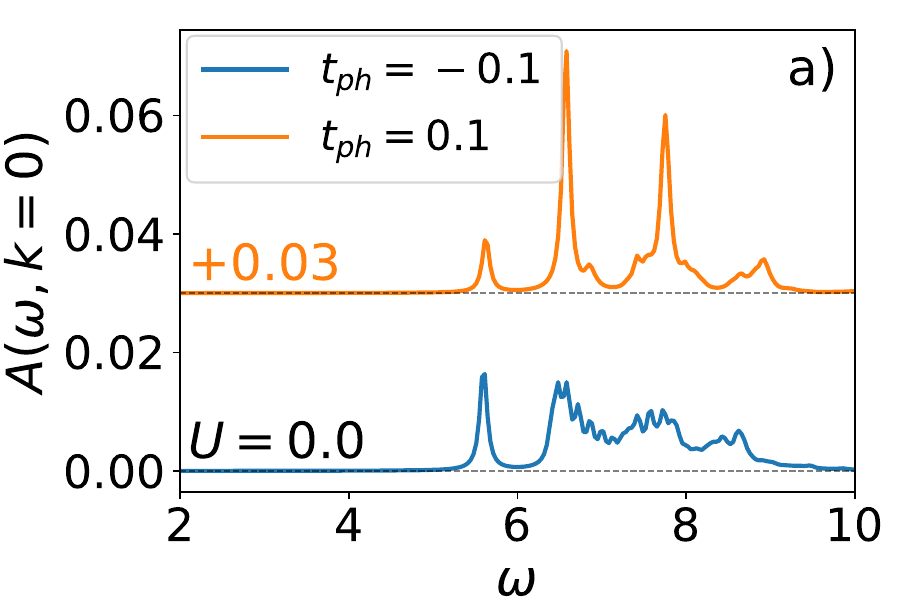} 
		\phantomsubcaption{}
		\label{fig6a}
	\end{subfigure}
	\begin{subfigure}[b]{0.5\columnwidth}
		\centering
		\includegraphics[width=1\linewidth]{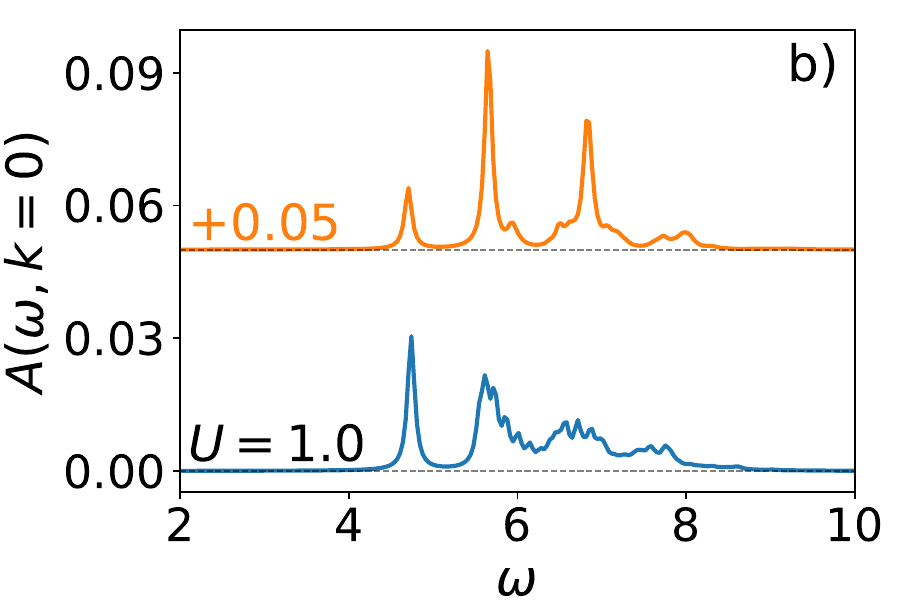} 
		\phantomsubcaption{}
		\label{fig6b}
	\end{subfigure}

    \vspace{-0.4cm}
    
    \begin{subfigure}[b]{0.5\columnwidth}
		\centering
		\includegraphics[width=1\linewidth]{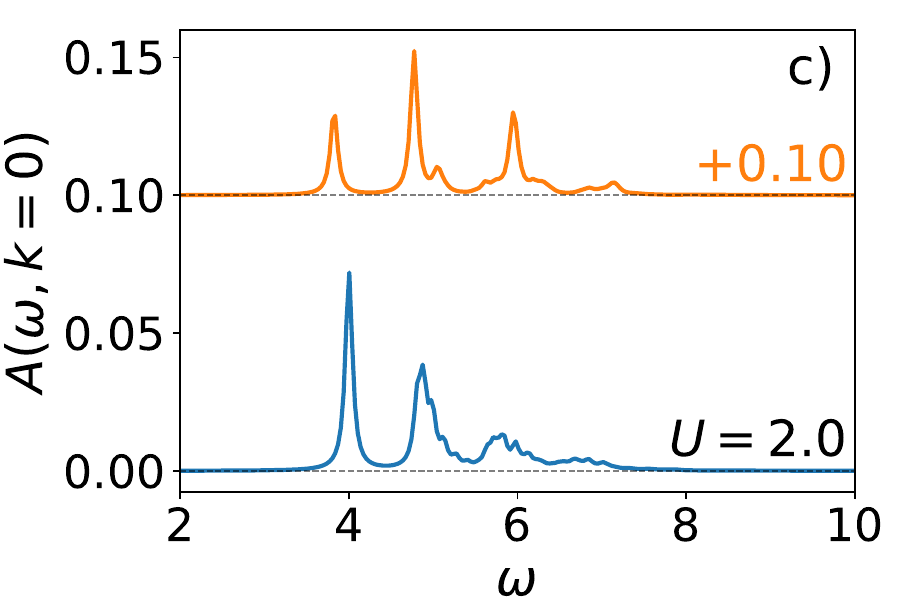} 
		\phantomsubcaption{}
		\label{fig6c}
	\end{subfigure}
	\begin{subfigure}[b]{0.5\columnwidth}
		\centering
		\includegraphics[width=1\linewidth]{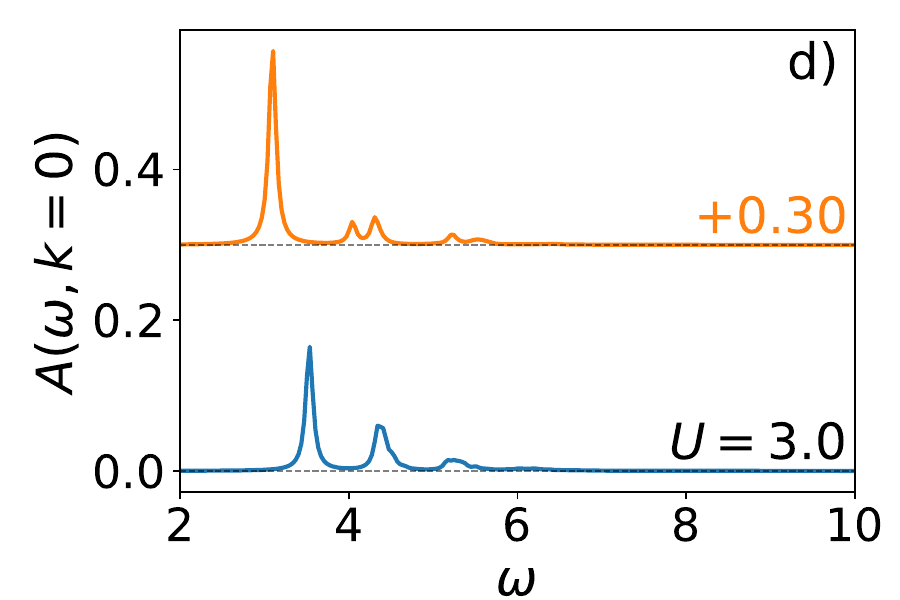} 
		\phantomsubcaption{}
		\label{fig6d}
	\end{subfigure}

    \vspace{-0.4cm}
    
	\captionsetup{justification=raggedright,singlelinecheck=false}
	\caption{The evolution of spectral functions $A(\omega,k = 0)$ with $U$ is presented for $t_{\mathrm{ph}} = {-0.1, 0.1}$. To display functions for both dispersions on the same figure, we vertically displaced them by the appropriate number, denoted above the displaced line in the same color. Artificial broadening was applied as usual.}
\label{fig6}
\end{figure}

\begin{figure}[!tbh]
	\includegraphics[width=0.9\columnwidth]{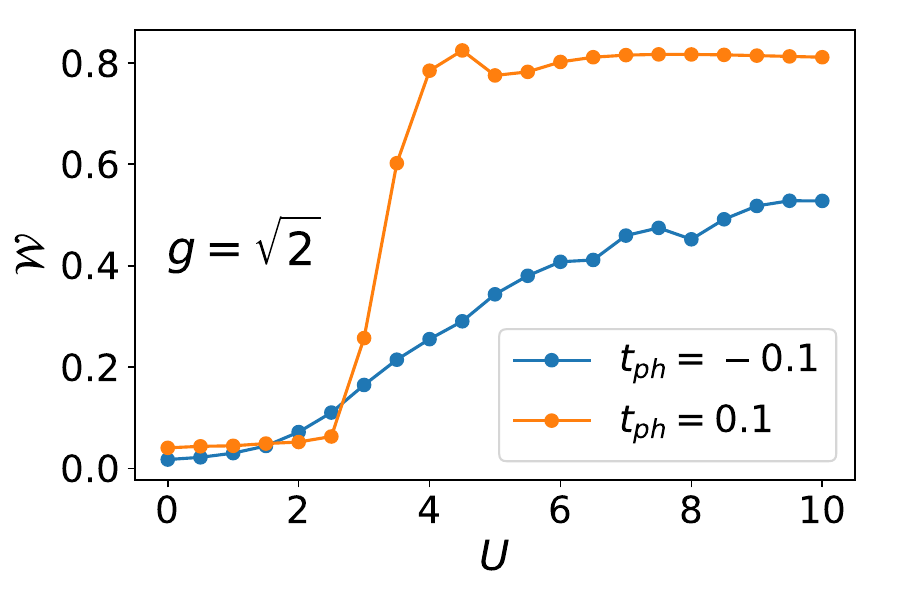}
	\captionsetup{justification=raggedright,singlelinecheck=false}
	\caption{The weight $\mathcal{W}$ as a function of $U$ for two different values of $t_{\mathrm{ph}}$ in a strongly EP coupled system ($g = \sqrt{2}$).}
	\label{fig7}
\end{figure}

Let's return our attention to Fig.~\ref{fig4}. When we neglect Coulomb repulsion between polarons, we observe a spectral function similar to the polaronic spectral function, a spectral function of electron addition in a previously empty system, in the regime of strong electron-phonon coupling \cite{Bonca2019, Bonca2021}. At the bottom of the spectra, there is a prominent polaronic band (represented with black dashed lines in Fig.~\ref{fig4} and labeled as P), corresponding to the ground state of a one-electron system with dispersion $\omega(K)=E_1^0(K) - E_2^0(K = 0)$, where $E_1^0(K)$ denotes the dispersion of the ground state of the one-electron system, and $E_2^0(K = 0)$ is the ground state energy at $K = 0$ of the two-electron system ground state. This band spreads throughout the entire Brillouin zone, separated from the upper bands, which are remnants of a continuum of states from the one-electron spectra. Their shape resembles that of the polaronic band. In Fig.~\ref{fig4}b, we observe a strong peak in the middle of the Brillouin zone, above the polaronic band but below the continuum. This peak corresponds to the bound state we observed in Fig.~\ref{fig1}f, thus we label it as B, and its dispersion, $\omega(K)=E_1^1(K) - E_2^0(K = 0)$, is represented with a blue dashed line in Fig.~\ref{fig4}b. It is visible just close to the center of the Brillouin zone before the state enters the continuum of one-electron states, as seen in Fig.~\ref{fig1}f. Interestingly, in a system with downward dispersion at $U = 0$, the spectral weight of this peak is greater than the spectral weight of the polaronic band. As shown in Fig.~\ref{fig6}, the spectral weight of the B-peak gradually diminishes with increasing $U$, transferring its weight to the P-peak. \\
Comparing the left and right columns of Fig.~\ref{fig4} along with Fig.~\ref{fig5}, several important observations can be made. Firstly, the P-band, along with the entire spectrum, gradually shifts to lower values of energy before converging towards $\omega(K\sim 0)=-E_1^0(K\sim 0)$ at some finite $U$. This is a consequence of the binding energy decreasing to zero with increasing $U$. At sufficiently large values of $U$, the binding of polarons becomes energetically unfavorable, and polarons behave as two undisturbed entities, as evidenced by the concentration of spectral weight in the P-band at $k = 0$. This conclusion can also be drawn by comparing spectral functions with the ones obtained for a polaron in the paper by Bonča and Trugman\cite{Bonca2022}. However, in our system due to finite system size, polarons still interact with each other, leading to the appearance of additional peaks at $k = \pm 2\pi/L$ next to the central peak, as observed in Figs.~\ref{fig4}c,\ref{fig4}d.  \\ Secondly, the spectral weight, which is spread throughout a large energy interval at small values of $U$, concentrates into a central peak of the P-band as $U$ increases. This is also depicted in Fig.~\ref{fig7}, which shows the weight $\mathcal{W}$ as a function of $U$ for two different values of $t_\mathrm{ph}$ in a strongly electron-phonon coupled system ($g = \sqrt{2}$). In this figure, we observe a jump in $\mathcal{W}$ around $U = 4$ for $t_{\mathrm{ph}} = 0.1$, which corresponds to a direct transition from the $S0$ bipolaron to an unbound pair of polarons. In contrast, for $t_{\mathrm{ph}} = -0.1$, $\mathcal{W}$ increases gradually. In this case, the bipolaron transitions through different types ($S0$, $S1$ and $S2$) before eventually unbinding, see also ref.~\cite{Kovac2024}. The increase in $\mathcal{W}$ with $U$ is due to a decrease of the bipolaron's binding energy as well as its effective mass. Weakly bound bipolaron as well as separated polarons are at large $U$ less localized in the real space and consequently more localized in  $k$-space at $k = 0$. \\ Thirdly, the spectral functions of a system with strong electron-electron repulsion exhibit multiphoton bands represented by colored dashed lines in Figs.~\ref{fig4}c and \ref{fig4}d. In addition to the single-phonon band, two- and three-phonon continuums are clearly visible. The single-phonon band, with a dispersion $\omega_{1\mathrm{ph}} = -E_1^0(K = 0) + \omega_0 + 2t_{\mathrm{ph}}\cos(q)$, is represented by yellow dashed lines. The two-phonon continuum is located in the interval given by $\omega^{\pm}_{2\mathrm{ph}} = -E_1^0(K = 0) + 2\omega_0 \pm 4t_\mathrm{ph}\cos(q/2)$, as indicated by the pink dashed lines. The three-phonon continuum appears between $\omega^-_{3ph} = -E_1^0(K = 0) + 3\omega_0 - 6t_\mathrm{ph}\cos((q\pm\pi)/3)$ and $\omega^+_{3ph} = -E_1^0(K = 0) + 3\omega_0 + 6t_\mathrm{ph}\cos(q/3)$ for $t_\mathrm{ph} > 0$. The vertical position of $\omega^+_{3\mathrm{ph}}$ and $\omega^-_{3\mathrm{ph}}$ is reversed for upward dispersion. The boundaries of the three-phonon continuum are indicated in the figures by white dashed lines. Not only does the shape of the phonon continuum vary, but also its spectral weight distribution depends on $t_{\mathrm{ph}}$. Comparing the two-phonon continuum of Figs.~\ref{fig4}c and \ref{fig4}d, we observe that for $t_{\mathrm{ph}} = -0.1$, the spectral weight is larger closer to the lower bound, whereas the opposite is true for $t_{\mathrm{ph}} = 0.1$. \\
The evolution of spectral function with $U$ can be better understood by examining the wavefunctions of a two-electron system in the limits of $U = 0$ and $U\rightarrow\infty$. For $U=0$, the ground state of a coupled bipolaron system without Coulomb repulsion can be expressed as $\ket{BP} \sim \big[ \sum_k \alpha_k c^\dagger_{k\uparrow} c^\dagger_{-k\downarrow} + \sum_{k,q} \alpha_{k,q} c^\dagger_{k\uparrow} c^\dagger_{-k-q,\downarrow} b^\dagger_q+\dots\big]|0\rangle$  , representing two localized, bound polarons. In contrast, when $U$ is sufficiently large, the system's ground state will consist of two independent polarons, which can be described by the product state $ |\psi^{(0,1)}_{0\uparrow}\rangle \otimes |\psi^{(0,1)}_{0\downarrow}\rangle$. At $U=0$, there is an overlap between the ground state of a polaron and the state obtained from $\ket{BP}$ by removing one electron throughout the Brillouin zone. On the other hand, as $U\rightarrow\infty$, the spectral function exhibits characteristics of the polaron spectrum, featuring a prominent peak at $k = 0$ and $\omega = -E_1^0(K=0)$, along with continua above it that correspond to different numbers of phonons. This is because the removal of one electron leaves the other polaron unaffected.

\begin{figure}[!tbh]
    \hspace*{-0.1\linewidth}
	\begin{subfigure}[b]{1\columnwidth}
		\centering
		\includegraphics[width=1.05\linewidth]{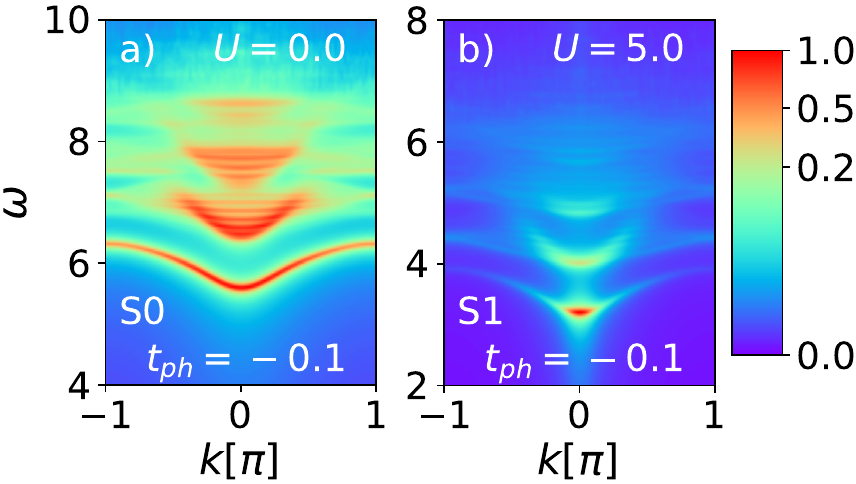} 
		\phantomsubcaption{}
		\label{fig8a}
	\end{subfigure}
	
    \vspace{-0.4cm}

    \begin{subfigure}[b]{1\columnwidth}
		\centering
		\includegraphics[width=1\linewidth]{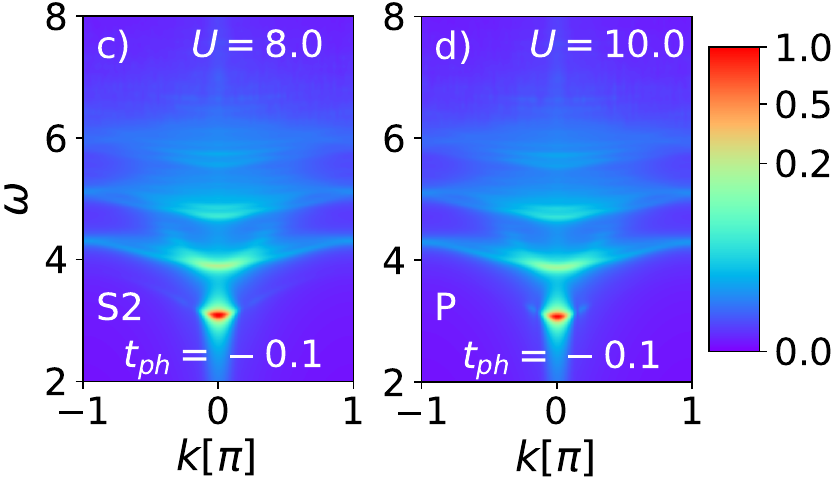} 
		\phantomsubcaption{}
		\label{fig8b}
	\end{subfigure}
	
    \vspace{-0.4cm}
 
	\captionsetup{justification=raggedright,singlelinecheck=false}
	\caption{The spectral functions $A(\omega,k)$ are presented for a system with $t_{\mathrm{ph}}$ and varying values of $U$, which determine whether a bipolaron of type S0, S1, or S2 is present, or polarons are unbound (P). An artificial broadening of delta functions is applied, determined by the parameter $\eta = 0.05$. The normalization procedure is elucidated in the text.}
\label{fig8}
\end{figure}

Fig.~\ref{fig8} illustrates the evolution of the spectral function of a system with $g = \sqrt{2}$ and upward phonon dispersion with increasing $U$. We selected values of $U$ at which different types of bipolarons are present in the system, denoted at the left bottom part of the figures.
In the S0 bipolaron regime (Fig.~\ref{fig8}a), the spectral function is spread throughout a large portion of the energy interval. The polaronic band exhibits significant spectral weight across the Brillouin zone. In addition,  close to $k = 0$, there is a notable contribution from excited levels of one-electron spectra. As $U$ increases, the system crosses over to the S1 bipolaron regime, the corresponding spectral function is shown in Fig.~\ref{fig8}b. Here, the polaronic band and the entire spectral function shift to lower energy values, approaching the lower limit of $\omega(K=0)=-E_1^0(K = 0)$ as $U \rightarrow \infty$. The spectral weight is visibly nonuniform across the Brillouin zone, with most of the spectral weight concentrated at $k = 0$. The P-band remains visible for all values of $k$ but gradually fades toward the edges of the Brillouin zone. In the continuum, traces of one-electron states with $k \neq 0$ persist, while the single-phonon band and two- and three-phonon continuum described earlier, as seen in Figs.~\ref{fig4}c, become clearly visible. Moving forward, in the S2 bipolaron regime (Fig.~\ref{fig8}c), the P-band remains are visible close to the center of the Brillouin zone. At sufficiently large $U$, polarons are no longer bound.  Such a case is shown in  Fig.~\ref{fig8}d where there is no polaronic band, only a strong peak emerges at $k = 0$, accompanied by two satellite peaks at $k = \pm 2\pi/L$  that are a consequence of the finite system size. 

\begin{figure}[!tbh]

        \begin{subfigure}[b]{1\columnwidth}
		\centering
		\includegraphics[width=1\linewidth]{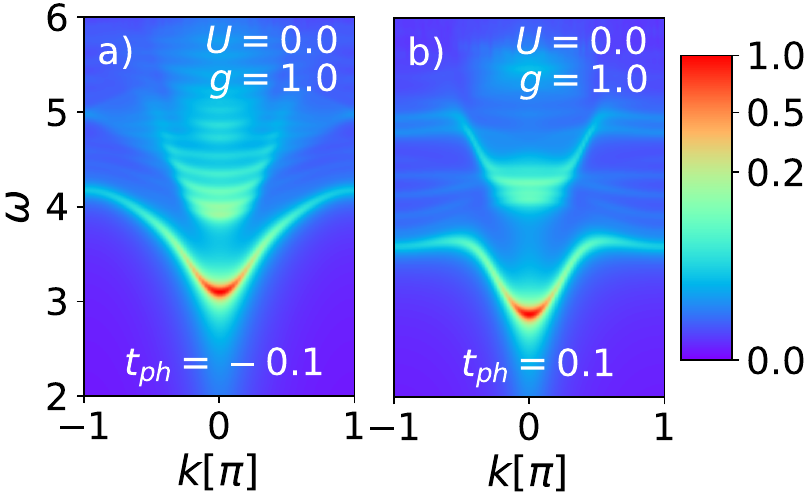} 
		\phantomsubcaption{}
		\label{fig9a}
	\end{subfigure}
 
	\begin{subfigure}[b]{1\columnwidth}
		\centering
		\includegraphics[width=1\linewidth]{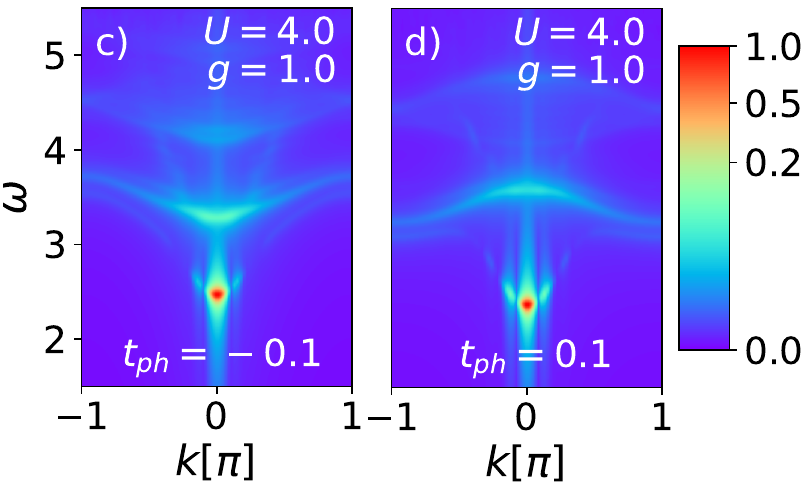} 
		\phantomsubcaption{}
		\label{fig9b}
	\end{subfigure}

	\captionsetup{justification=raggedright,singlelinecheck=false}
	\caption{The spectral functions $A(\omega,k)$ of a system in an intermediate electron-phonon coupling regime ($g = 1$) are presented. Each column corresponds to a different phonon dispersion, while each row represents a different value of $U$. The top row depicts the system without Coulomb interaction, while the bottom one corresponds to a scenario where $U$ is sufficient to prevent bipolaron formation. An artificial broadening of delta functions is applied, determined by the parameter $\eta = 0.05$. The normalization procedure is explained in the text.}
\label{fig9}
\end{figure}

Fig.~\ref{fig9} presents spectral functions of the intermediate EP coupling regime ($g = 1$) with $t_{\mathrm{ph}}\pm 0.1$ and two values of $U=0, 4$, to facilitate a comparison with Fig.~\ref{fig4}. Firstly, comparing the figures with $U = 0$, we observe that for weaker EP coupling, the contribution of the polaronic band to the spectral weight becomes larger, and excited states of the one-electron system play a less important role - there is much less intensity at higher $\omega$ than in Figs.~\ref{fig4}a and b. The bandwidth is larger at smaller  $g$, consistent with a smaller effective mass. 
In case of larger $U = 4$ where the system is composed of two separate polarons, we observe  a smaller overall downward shift of the  spectral function, a consequence of weaker binding energy at $U=0$ in comparison to $g=\sqrt{2}$ case. The single phonon band and multiphonon  continuum are less visible in the case of weaker electron-phonon coupling and there is more spectral weight concentrated around $k=0$. 
 
\begin{figure}[!tbh]
    
        \begin{subfigure}[b]{0.8\columnwidth}
		\centering
		\includegraphics[width=1\linewidth]{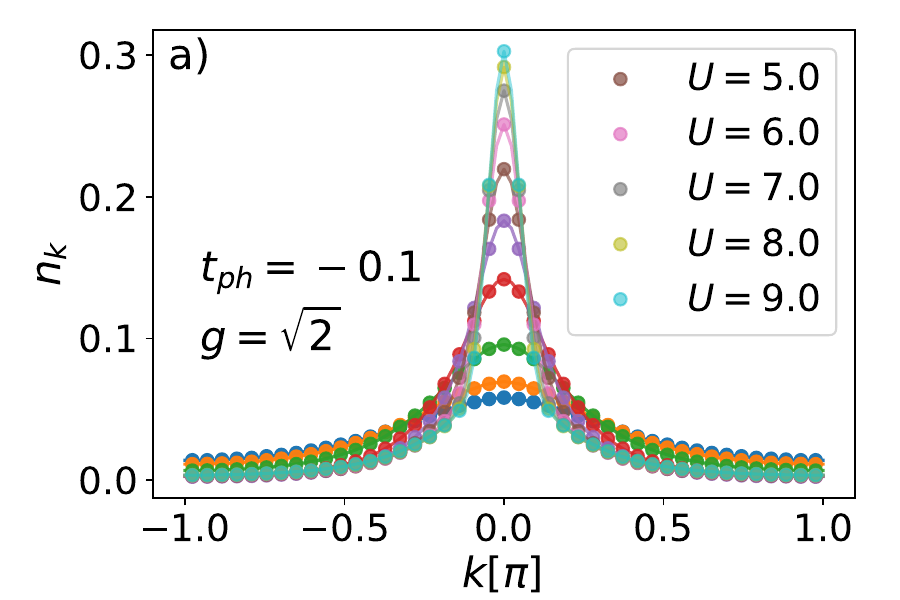} 
		\phantomsubcaption{}
		\label{fig10a}
	\end{subfigure}

	\begin{subfigure}[b]{0.8\columnwidth}
		\centering
		\includegraphics[width=1\linewidth]{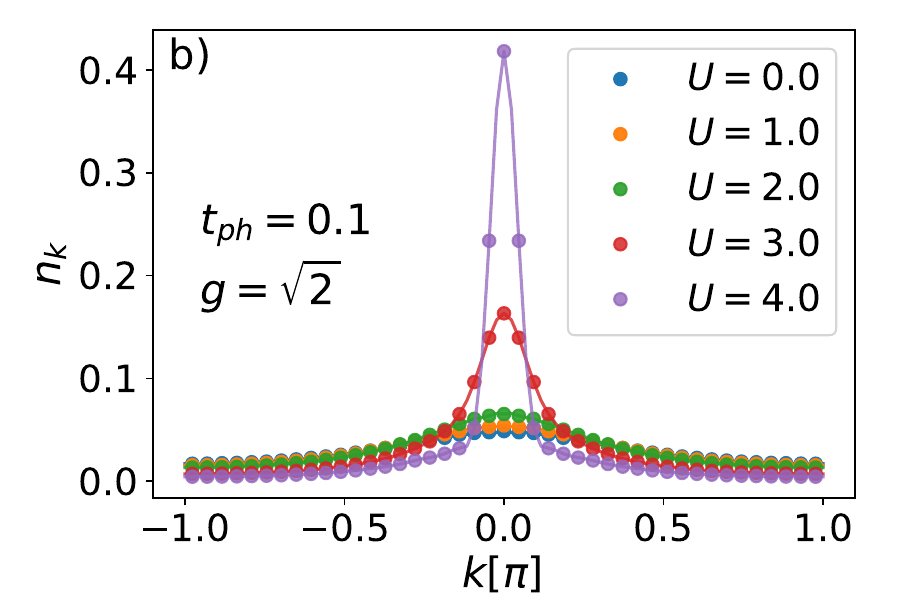} 
		\phantomsubcaption{}
		\label{fig10b}
	\end{subfigure}

	\captionsetup{justification=raggedright,singlelinecheck=false}
	\caption{Variation of the expectation value of the number operator with respect to wavevector $k$ under different conditions of phonon band: (a) upward phonon dispersion and (b) downward phonon dispersion, as a function of $U$. To enhance
    visualization, only every 2th data point is shown on the plot.}
\label{fig10}
\end{figure}

We now turn to sum rules. Integrating over the frequency interval according to Eq.~\eqref{SumRule}, we obtain  $n_{ks}$,  depicted in Fig.~\ref{fig10} as a function of $U$ for both phonon dispersion scenarios. This quantity provides additional insight into the properties of bipolaron augmenting those revealed by the binding energy and spectral functions. 
In systems dominated by the strong EP coupling, polarons form tightly bound bipolarons with large effective mass in the absence of $U$. 
Consequently, $n_{ks}$ exhibits a broader distribution across $k$ since due to a large effective mass the bipolaron is nearly localized. 
As $U$ increases, the bipolaron expands resulting in reduced effective mass\cite{Bonca2000}.  Eventually, when $U$ exceeds a critical threshold for polarons to form bipolaron, $n_{ks}$ becomes sharply peaked around $k = 0$. Fig.~\ref{fig10b} shows a sharp transition in \( n_{ks} \), shifting from a broad distribution to a sharply peaked function as \( U \) increases from 2 to 4. This change reflects the rapid crossover from a bipolaron state to two separate polarons for \( t_{ph} = 0.1  \), consistent with our previous results~\cite{Kovac2024}. In contrast, when \( t_{ph} = -0.1 \), the transition from a bipolaron to an unbound state is more gradual, as depicted in Fig.~\ref{fig10a}. In this case, the bipolaron passes through several intermediate phases (e.g., \( S0 \), \( S1 \)) before evolving into two quasi-independent, weakly interacting particles (due to the finite system size). The evolution of \( n_{ks} \) continues in a similar manner, with the polarons ultimately becoming unbound around \( U \approx 9 \).  
\\
To assess the accuracy of our spectral function, we calculate the sum rule, see Eqs.~\eqref{SumRule} and \eqref{SumRule1}, by summing up $n_{ks}$ over a discrete $k$ interval used in the computation of $A(\omega,k)$. Our sum rule aligns well with analytical predictions, although deviations are more pronounced for smaller $U$ values where the spectral function spans a broader $\omega$ interval. 


\section{Conclusions}

Using an efficient variational method, we computed the spectral function of a two-electron system within the framework of the Holstein-Hubbard model with dispersive quantum optical phonons. Our primary focus was on examining the interplay between phonon dispersion and Coulomb repulsion and their effects on the spectral function.\\
We observed a detectable effect of phonon dispersion on spectral functions. In regimes where EP coupling dominates over Coulomb interaction, the spectral function spreads throughout a large portion of the frequency interval, with spectral weight distribution strongly influenced by the phonon dispersion. Specifically, in systems with upward phonon dispersion, the quasiparticle band carries most of the spectral weight. Conversely, in systems with downward dispersion at $U=0$, the spectral weight is in the strong EP coupling regime primarily carried by the bound state that is separated from the continuum of states around the center of the Brillouin zone.  The spectral function in this regime is markedly different from the polaronic spectral function, reflecting the distinct nature of the bipolaronic wavefunction compared to that of two separate polarons. As $U$ increases, the bipolaron becomes unstable and separates into two polarons. Consequently, polarons behave as separate entities with repulsive interaction, while the spectral function resembles that of isolated polarons. \\

Building on our previous research \cite{Kovac2024}, we aimed to identify differences between spectral functions corresponding to different bipolaronic regimes identified as $S0, S1$, and $S2$. While visible differences are evident in Fig.~\ref{fig8}, we believe they may not be sufficient to conclusively determine the type of bipolaron solely based on the spectral function.


\begin{acknowledgments}
	J.B. and K.K. acknowledge the support by program No. P1-0044 of the Slovenian Research Agency (ARIS). J.B. acknowledges discussions with M. Berciu, S.A. Trugman, A. Saxena and support from the Center for Integrated Nanotechnologies, a U.S. Department of Energy, Office of Basic Energy Sciences user facility and Physics of Condensed Matter and Complex Systems Group (T-4) at Los Alamos National Laboratory. 
\end{acknowledgments} 

\appendix

\section*{Appendix A: Finite system size analysis}

\begin{figure}[!tbh]
	\begin{subfigure}[b]{0.5\columnwidth}
		\centering
		\includegraphics[width=1\linewidth]{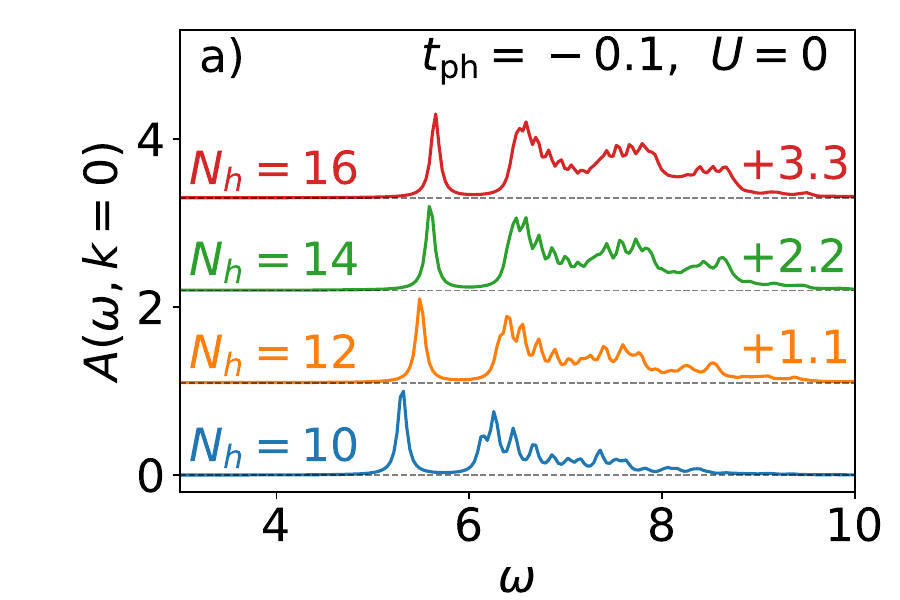} 
		\phantomsubcaption{}
		\label{figA1a}
	\end{subfigure}
	\begin{subfigure}[b]{0.5\columnwidth}
		\centering
		\includegraphics[width=1\linewidth]{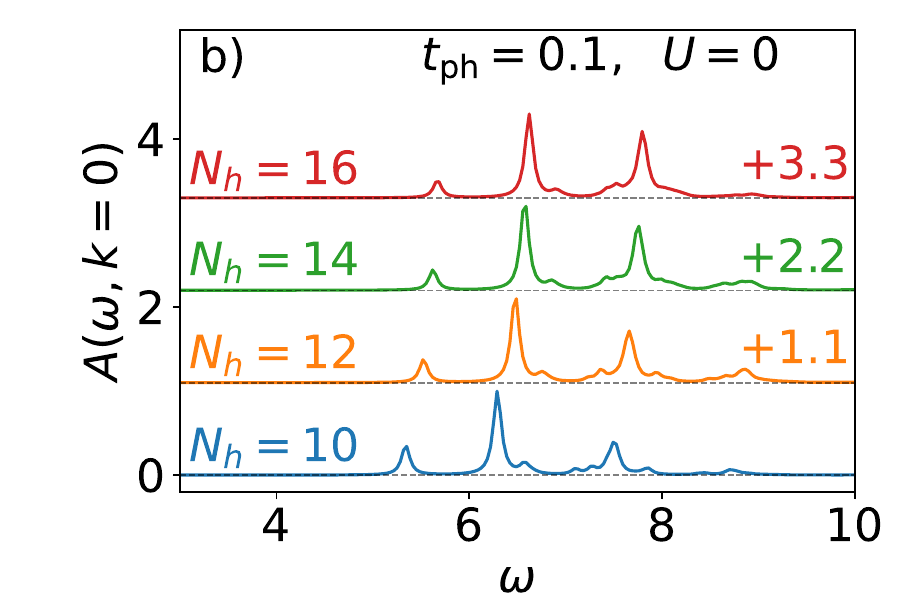} 
		\phantomsubcaption{}
		\label{figA1b}
	\end{subfigure}

    \vspace{-0.4cm}
    
    \begin{subfigure}[b]{0.5\columnwidth}
		\centering
		\includegraphics[width=1\linewidth]{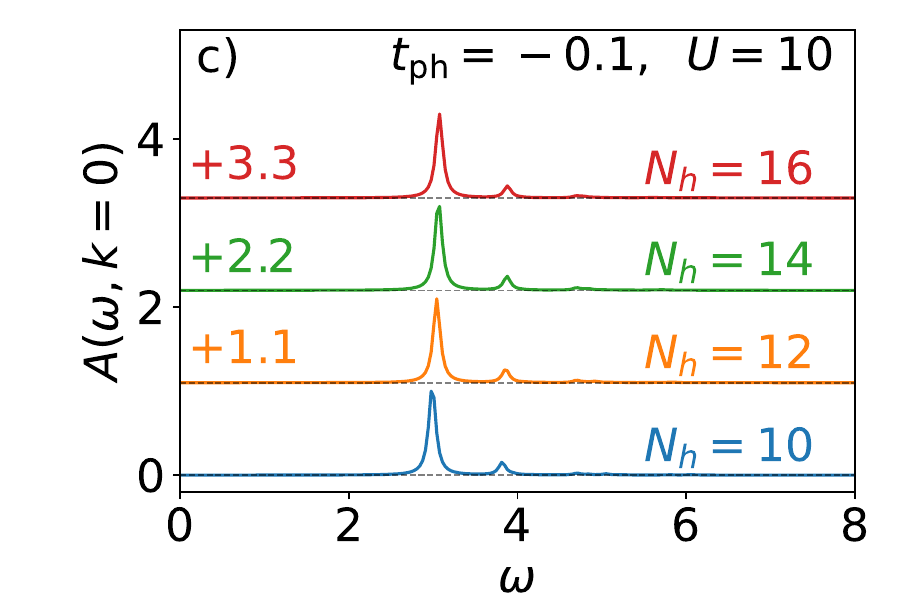} 
		\phantomsubcaption{}
		\label{figA1c}
	\end{subfigure}
	\begin{subfigure}[b]{0.5\columnwidth}
		\centering
		\includegraphics[width=1\linewidth]{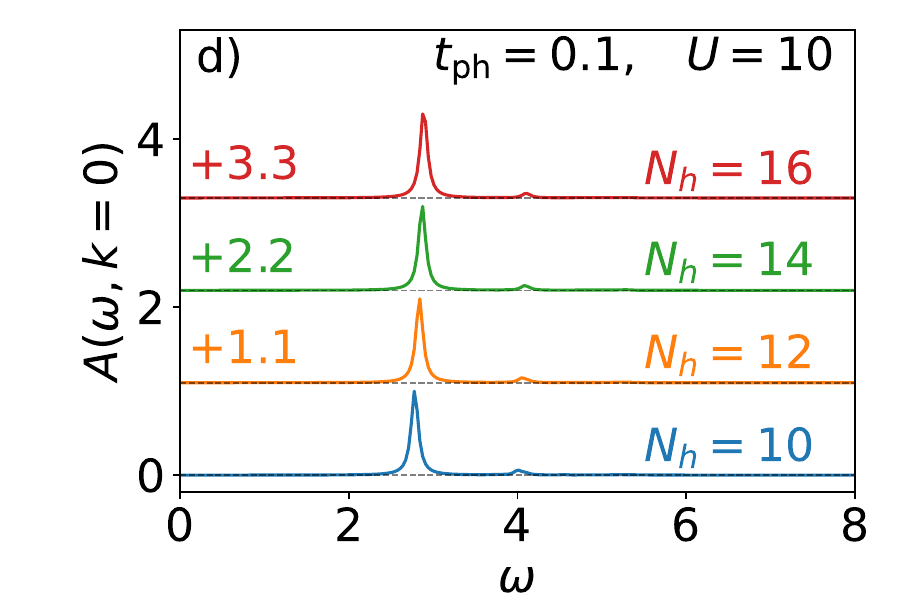} 
		\phantomsubcaption{}
		\label{figA1d}
	\end{subfigure}

    \vspace{-0.4cm}
    
	\captionsetup{justification=raggedright,singlelinecheck=false}
	\caption{The evolution of spectral functions $A(\omega,k = 0)$ with system size determined by $N_h$ is presented for $t_{\mathrm{ph}} = {-0.1, 0.1}$ and $U = {0, 10}$. To display functions for different values of $N_h$ on the same figure, we vertically displaced them by the appropriate number, denoted above the displaced line in the same color. Artificial broadening was applied as usual. The normalization procedure is
    explained in the text. }
\label{figA1}
\end{figure} 

By comparing spectral functions at different values of $N_h$, which determine not only the size of our system but also the maximum number of phonon excitations present at a specific site, we aim to demonstrate that our method is sufficient to determine the spectral function of a dilute system in the thermodynamic limit. For easier analysis, we focus on a slice of the spectral function with $k = 0$. \\
Fig.~\ref{figA1} shows the evolution of the spectral function $\smash{A(\omega,k = 0)}$ with system size for \(t_{\mathrm{ph}} = {-0.1, 0.1}\) and $\smash{U = {0, 10}}$. We observe small differences at large values of \(U\), while more pronounced variations occur in a system without Coulomb repulsion. In the absence of \(U\), the spectral function is spread over a larger portion of the energy interval, with system size influencing its fine structure within the continuum region. Meanwhile, the shapes of the \(P\)- and \(B\)-peaks remain relatively stable. The changes are more pronounced between plots at lower values of \(N_h\), indicating the convergence of the spectral function with increasing \(N_h\).
\begin{figure}[!tbh]
	\begin{subfigure}[b]{0.5\columnwidth}
		\centering
		\includegraphics[width=1\linewidth]{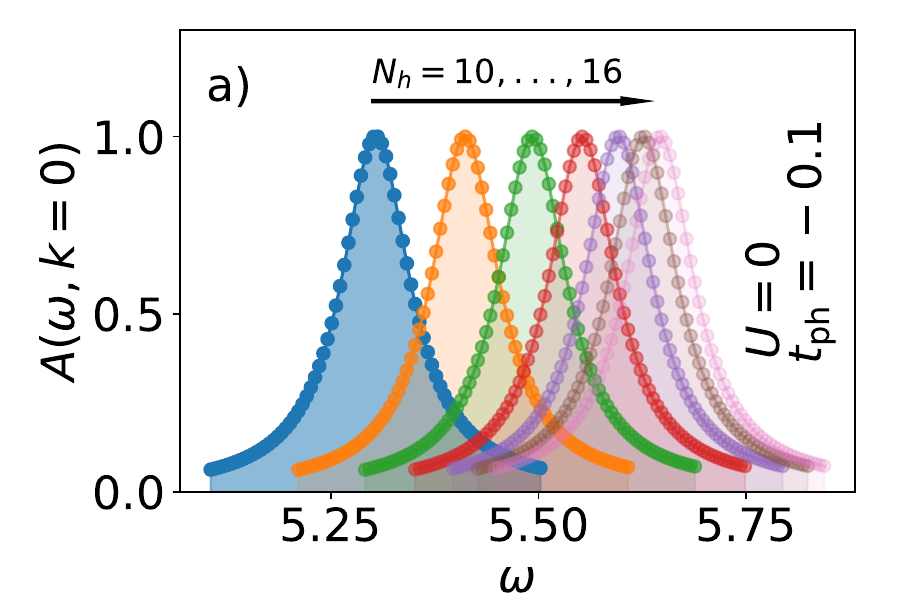} 
		\phantomsubcaption{}
		\label{figA2a}
	\end{subfigure}
	\begin{subfigure}[b]{0.5\columnwidth}
		\centering
		\includegraphics[width=1\linewidth]{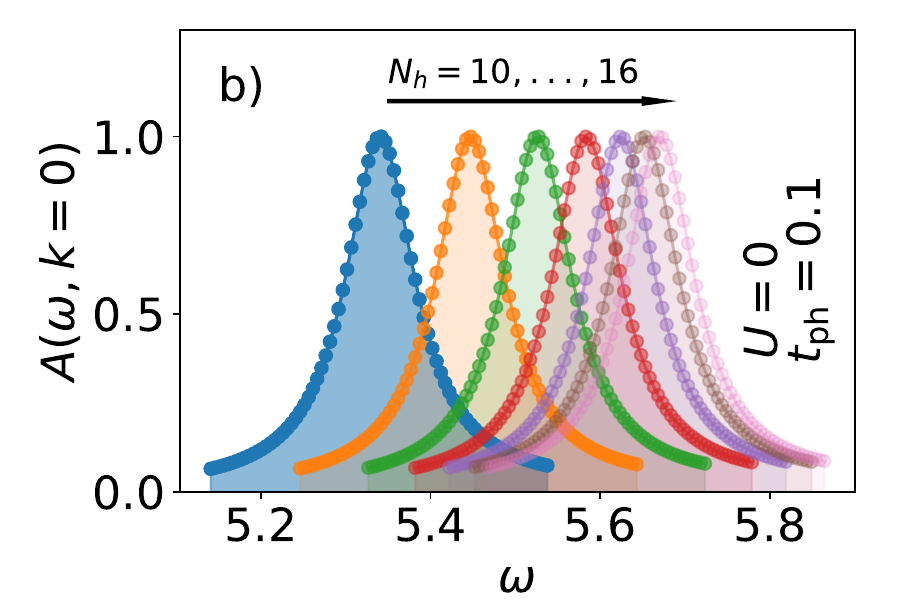} 
		\phantomsubcaption{}
		\label{figA2b}
	\end{subfigure}

    \vspace{-0.4cm}
    
    \begin{subfigure}[b]{0.5\columnwidth}
		\centering
		\includegraphics[width=1\linewidth]{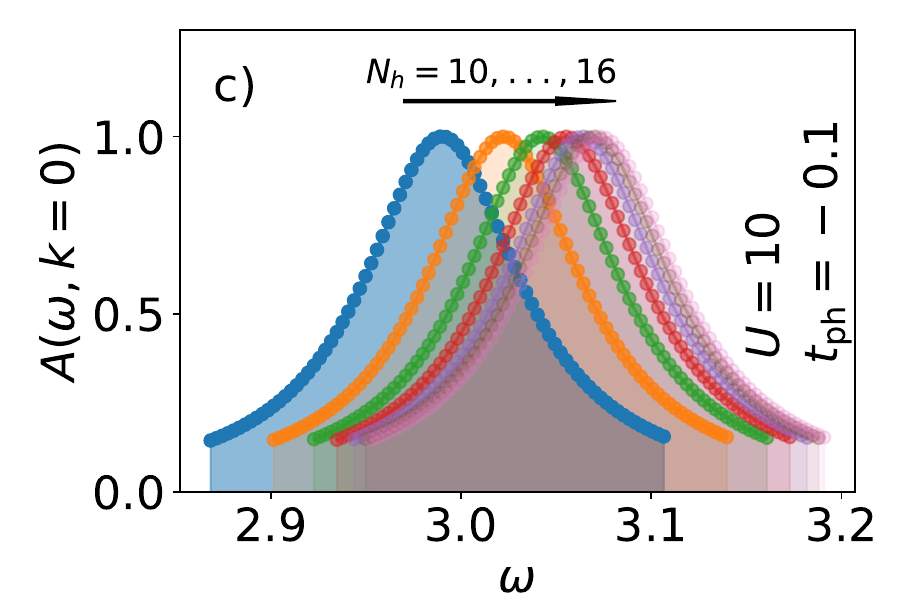} 
		\phantomsubcaption{}
		\label{figA2c}
	\end{subfigure}
	\begin{subfigure}[b]{0.5\columnwidth}
		\centering
		\includegraphics[width=1\linewidth]{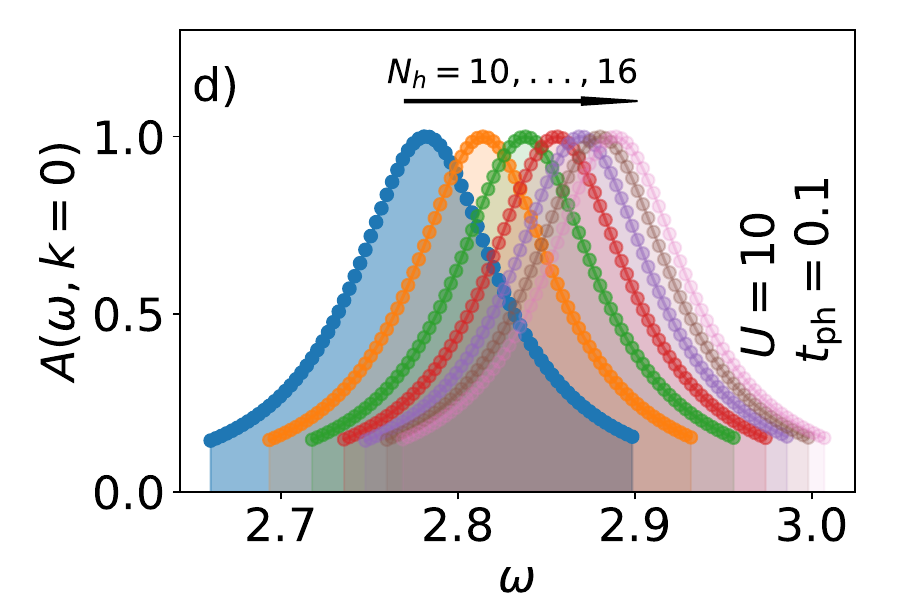} 
		\phantomsubcaption{}
		\label{figA2d}
	\end{subfigure}

    \vspace{-0.4cm}

    \begin{subfigure}[b]{0.5\columnwidth}
		\centering
		\includegraphics[width=1\linewidth]{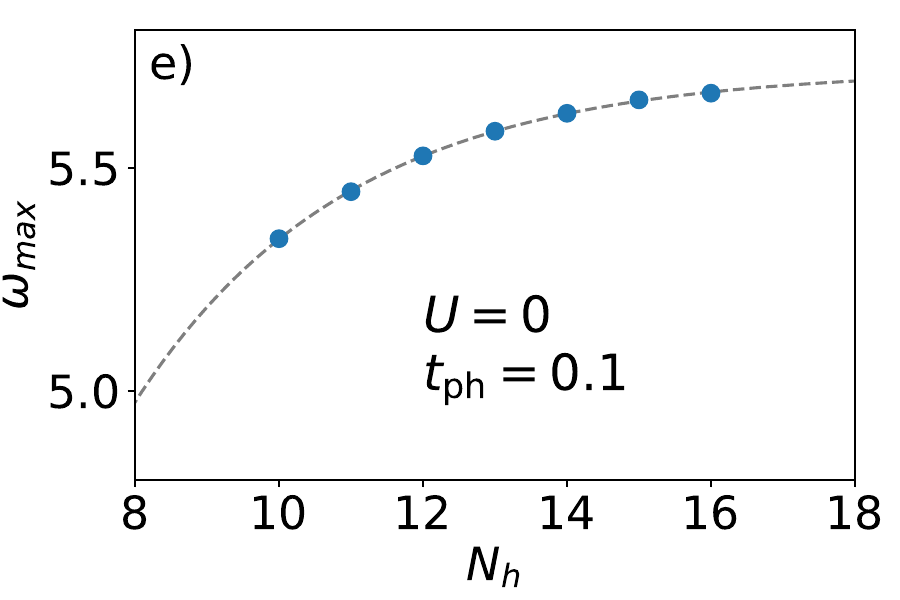} 
		\phantomsubcaption{}
		\label{figA2e}
	\end{subfigure}
	\begin{subfigure}[b]{0.5\columnwidth}
		\centering
		\includegraphics[width=1\linewidth]{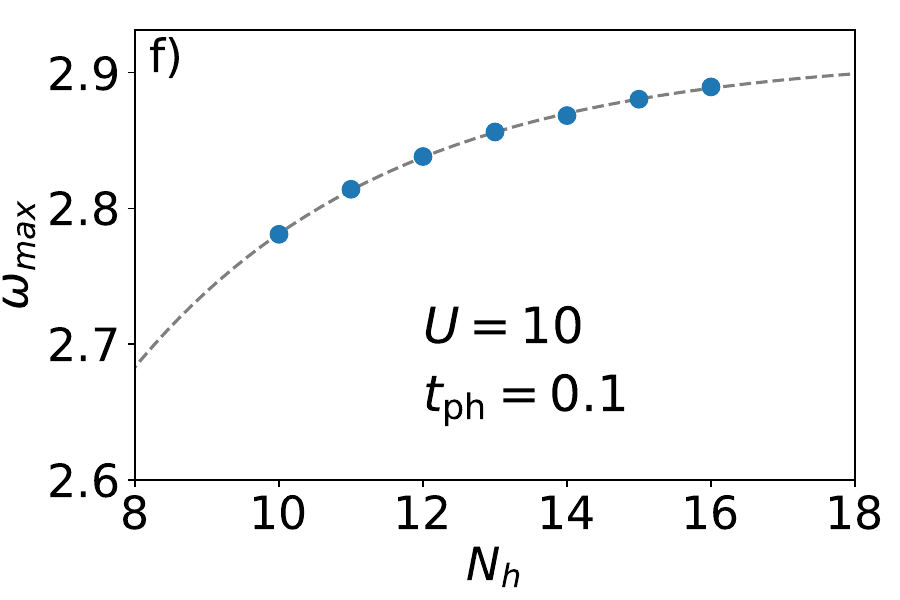} 
		\phantomsubcaption{}
		\label{figA2f}
	\end{subfigure}

    \vspace{-0.4cm}
    
	\captionsetup{justification=raggedright,singlelinecheck=false}
	\caption{The drifting of the highest peak of the spectral functions $A(\omega,k = 0)$, calculated for increasing values of $N_h$, is indicated by the black arrow. This is shown in Figures a to d for $t_{\mathrm{ph}} = {-0.1, 0.1}$ and $U = {0, 10}$, where artificial broadening and the normalization procedure were applied as usual. Figs.~e) and f) depict the drift of the peak position $\omega_{max}$ as a function of $N_h$ (blue dots), along with the fitted exponential function as explained in the text (grey dashed line).}
\label{figA2}
\end{figure} 

\begin{figure*}[!tbh]
	\begin{subfigure}[b]{0.7\columnwidth}
		\centering
		\includegraphics[width=1\linewidth]{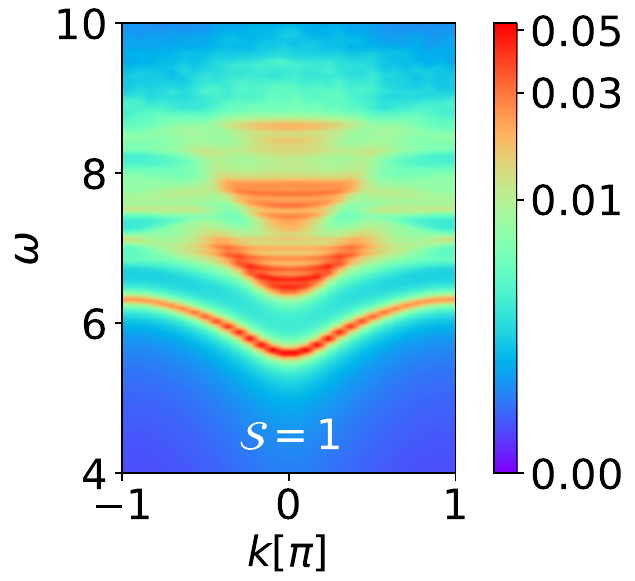} 
		\phantomsubcaption{}
		\label{figB1a}
	\end{subfigure}
	\begin{subfigure}[b]{0.7\columnwidth}
		\centering
		\includegraphics[width=1\linewidth]{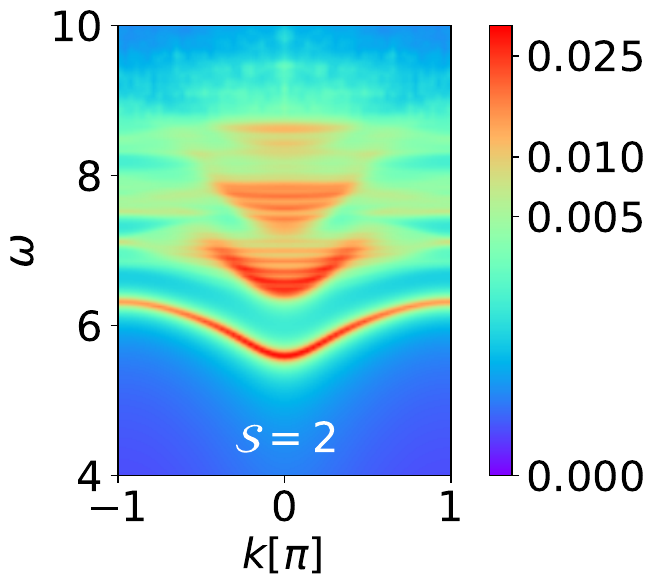} 
		\phantomsubcaption{}
		\label{figB1b}
	\end{subfigure}
    \begin{subfigure}[b]{0.7\columnwidth}
		\centering
		\includegraphics[width=1\linewidth]{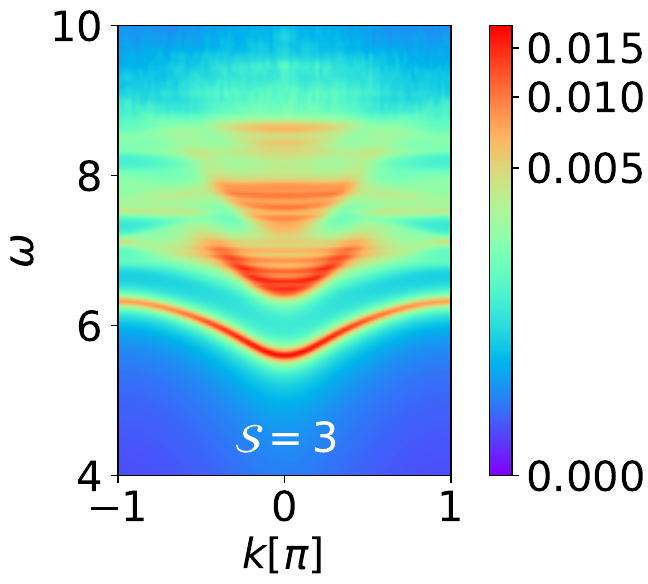} 
		\phantomsubcaption{}
		\label{figB1c}
	\end{subfigure}

    \vspace{-0.4cm}
    
	\captionsetup{justification=raggedright,singlelinecheck=false}
	\caption{Spectral function $A(\omega,k)$ calculated for increasing values of $\mathcal{S} = 1,2,$ and $3$ with parameters $U = 0.0$, $t_{ph} = -0.1$ and $g = \sqrt{2}$. Artificial broadening was applied using $\eta = 0.05$.}
\label{figB1}
\end{figure*}
In addition to minor changes in shape, we also observe a drifting of peaks to higher values of \(\omega\) at both \(U = 0\) and \(U = 10\). This feature is highlighted in Fig.~\ref{figA2}. The drifting at \(U = 0\) corresponds to the limitation in the number of phonon excitations. At small values of \(N_h\), this number is insufficient to produce the correct phonon cloud, resulting in a higher energy of the bipolaron state compared to the thermodynamic limit. On the other hand, the drifting at \(U = 10\) has a different physical origin: it is a consequence of polarons being closer to each other than they would prefer, leading to a higher energy of the two-polaron state due to increased electron-electron repulsion.\\
In both cases, as \(N_h\) increases, the shifting becomes very small, indicating that the conclusions drawn in our paper can be extended to the thermodynamic limit. To quantify this, we present exponential fit of the form $a\exp{bx}+c$ to the peak positions as shown in Figs.~\ref{figA2}e and \ref{figA2}f. The fitting error is small, and the value to which the exponential function converges is close to the value at $N_h = 16$, affirming the stability of the results. For instance, the quasiparticle peak for $U = 0, 10$ and $t_\mathrm{ph} = 0.1$, as shown in Figs.~\ref{figA2}e and \ref{figA2}f,  converges to $[5.7215, 2.9135]$ with a fitting error $\sim0.1\%$, while for $N_h = 16$ respective peak values are $[5.6683, 2.8894]$.

    


\section*{Appendix B: Scaling factor} \label{appendixB}

For a finite-sized system, the allowed values of $k$-vectors at which the spectral function can be calculated can be increased by forming a ring from the chain and exposing it to a magnetic flux. This adjustment stems from the modification of the free electron energy dispersion in such a system, expressed as $\varepsilon(k,\theta)=-2t_{el} \cos(k+\theta)$, where $\theta$ represents the Peierls phase due to magnetic flux through the ring. By appropriately selecting $\theta$, the energy dispersion, and consequently the spectral function, of the finite system becomes smoother, mimicking the behavior of an infinite system. \\
In numerical calculations, this effect is simulated by increasing the number of $k$-points at which the spectral function is evaluated, scaled by a factor $\mathcal{S}$ (referred to as the scaling factor). This manipulation, however, must preserve the sum rule, which states that the summation of the spectral function over the entire $\omega$ and $k$ space equals 1, corresponding to the number of electrons with a specific spin. Since the number of $k$-points is increased by the scaling factor, the spectral function must be rescaled accordingly to maintain the sum rule: $A(\omega, k) \longrightarrow A(\omega, k)/\mathcal{S}$. \\
Figure~\ref{figB1} illustrates the effect of the scaling factor $\mathcal{S}$ on the spectral function. The parameters used are $N_h =16$, $t_\mathrm{el} =1$, $\omega_0 = 1$, $t_\mathrm{ph} = -0.1$ and $g = \sqrt{2}$ with increasing $\mathcal{S}$. We observe that the overall structure of the spectral function remains unaffected by the choice of $\mathcal{S}$. The primary purpose of introducing $\mathcal{S}$ is to enhance the resolution of the spectral function: for $\mathcal{S} = 1$, the spectral function appears grainy, whereas for $\mathcal{S} = 3$, it appears significantly more continuous. \\
As seen from the values indicated on the color bars, the spectral function must be rescaled to satisfy the sum rule. However, this scaling does not alter the overall distribution of the spectral function across $k$ and $\omega$. This consideration also motivated us to plot normalized spectral functions, as the important features of the spectral function are more effectively displayed in this manner.

\section*{Appendix C: Changing phonon frequency}

So far we set the phonon frequency $\omega_0$ equal to the electron hopping parameter $t_\mathrm{el}$, both normalized to 1. This choice was made to focus on the effects of the Coulomb repulsion $U$  on the system's properties, particularly on its spectral function.  Even though large changes of $\omega_0$ from $\omega_0/t_\mathrm{el}\gg 1$ to $\omega_0/t_\mathrm{el}\ll 1$ shift the system from the anti--adiabatic to the adiabatic regime, we restrict our calculation in this Appendix to only small changes of $\omega_0$ due to restrictions of our numerical approach. 

It is important to note, however, that changing
$\omega_0$ while keeping the electron-phonon (EP) coupling $g$ constant alters the effective EP coupling strength $\lambda\sim g^2/2t_\mathrm{el}\sqrt{\omega_0^2-4t_\mathrm{ph}^2}$. Such a change could shift the system from one interaction regime to another, where the behavior differs significantly.  In the spectral function, this would manifest as a shift of the entire spectrum and a reduction of the quasiparticle bandwidth, as the effective hopping is modified by the EP interaction. 

\begin{figure}[!tbh]
	\includegraphics[width=1\columnwidth]{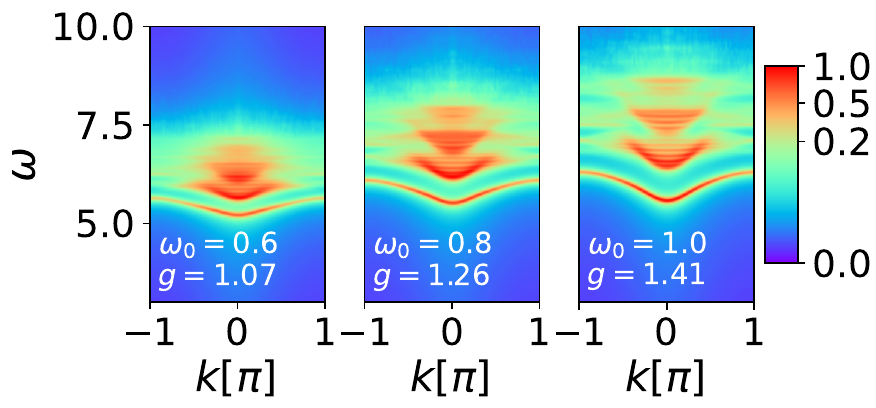}
	\captionsetup{justification=raggedright,singlelinecheck=false}
	\caption{The spectral function $A(\omega, k)$ is calculated for increasing values of $\omega_0 = 0.6, 0.8, 1.0$, with corresponding coupling constants $g = 1.07, 1.26, 1.41$, ensuring that the effective EP coupling strength remains constant at $\lambda \approx 1.0$. The parameters are set to $t_\mathrm{el} = 1$, $U = 0.0$, and $t_\mathrm{ph} = -0.1$. Artificial broadening with $\eta = 0.05$ and a normalization procedure were applied.}
	\label{figc1}
\end{figure}

The primary goal of our research is to analyze the effect of phonon dispersion on the spectral properties of bipolarons in the intermediate and strong coupling regimes. To ensure that the system remains within the same coupling regime while studying the influence of $\omega_0$ on the spectral function, we have changed  $g$ in tandem with $\omega_0$ so that $\lambda$ remains constant. Figure~\ref{figc1} illustrates how the value of $\omega_0$ affects the spectral properties of a system with negligible Coulomb repulsion and upward phononic dispersion. The EP coupling constant $g$ was adjusted to maintain a constant effective electron-phonon coupling strength. As $\omega_0$ decreases (from the right to the left figure), the spectrum becomes compressed, with reduced separation between energy regions corresponding to states with different phonon numbers. Additionally, the overlap between these regions increases. \\

For $\omega_0 = 0.6$, we observe a lack of spectral weight in the high-$\omega$ regime, i.e.  for $\omega\gtrsim 7.5$ and a small overall shift of spectrum, which is a numerical artifact - a consequence of the lack of sufficient phonon degrees of freedom in the variational space.  As $\omega_0$ decreases, the energy required to excite phonons in the system diminishes, a trend further amplified by the presence of phononic dispersion. For $t_\mathrm{ph} = -0.1$, the dispersion reduces the separation between the quasiparticle band and the continuum, particularly at the center of the Brillouin zone, where the phonon dispersion is given by $\omega_\mathrm{ph}(0) = \omega_0 + 2t_\mathrm{ph}$. \\


\newpage
\bibliography{manuaomm}
\end{document}